# Universal Polarization Transformations: Spatial programming of polarization scattering matrices using a deep learning-designed diffractive polarization transformer

Yuhang Li[1,2,3†], Jingxi Li[1,2,3†], Yifan Zhao[1,3], Tianyi Gan[1,3], Jingtian Hu[1,2,3], Mona Jarrahi[1,3] and Aydogan Ozcan[1,2,3*]

[1]Electrical and Computer Engineering Department, University of California, Los Angeles, CA, 90095, USA

[2]Bioengineering Department, University of California, Los Angeles, CA, 90095, USA

[3]California NanoSystems Institute (CNSI), University of California, Los Angeles, CA, 90095, USA

[†]These authors contributed equally to the work

[*]Correspondence to: ozcan@ucla.edu

**Keywords:** Polarization, Polarization transformations, Transformation optics, Optical computing, Optical neural networks

## Abstract

Spatial manipulation of polarization, with the controlled synthesis of optical fields having non-uniform polarization distributions, presents a challenging task. Here, we demonstrate universal polarization transformers based on an engineered diffractive volume, which can synthesize a large set of arbitrarily-selected, complex-valued polarization scattering matrices between the polarization states at different positions within its input and output field-of-views (FOVs). This framework comprises 2D arrays of linear polarizers with diverse angles, which are positioned between isotropic diffractive layers, each containing tens of thousands of diffractive features with optimizable transmission coefficients. We demonstrate that, after its deep learning-based training, this diffractive polarization transformer could successfully implement $N_i N_o$=10,000 different spatially-encoded polarization scattering matrices with negligible error within a single diffractive volume, where $N_i$ and $N_o$ represent the number of pixels in the input and output FOVs, respectively. We experimentally validated this universal polarization transformation framework in the terahertz part of the spectrum by fabricating wire-grid polarizers and integrating them with 3D-printed diffractive layers to form a physical polarization transformer operating at $\lambda$ = 0.75 mm wavelength. Through this set-up, we demonstrated an all-optical polarization permutation operation of spatially-varying polarization fields, and simultaneously implemented distinct spatially-encoded polarization scattering matrices between the input and output FOVs of a compact diffractive processor that axially spans 200×λ. This framework opens up new avenues for developing novel optical devices for universal polarization control, and may find various applications in, e.g., remote sensing, medical imaging, security, material inspection and machine vision.

## I. INTRODUCTION

Polarization of light indicates the geometrical orientation of electromagnetic wave oscillations, with its practical applications permeating various fields such as telecommunications[1–3], imaging[4–8], sensing[9–14], computing[15], and display technologies[16–18], among others. In pursuit of harnessing the full potential of this unique characteristic of light, there have been various efforts to achieve a higher level of flexibility and precision in controlling polarization. Traditional polarization modulation devices, such as polarizers and waveplates, have been widely employed for the manipulation of polarized beams or optical fields by implementing *spatially homogeneous* polarization transformations between an input and output. However, recent advances in micro- and nano-fabrication technologies have ushered in the development of more powerful polarization modulation elements using e.g., spatial light modulators (SLMs)[19,20] and metasurfaces[21–23], which enabled the creation of spatially structured polarization modulations. By meticulously designing the intricate structure responsible for light-matter interactions, one can modulate the polarized input optical field with spatially heterogeneous polarization transformations, subsequently converting the input polarization states into the desired output polarization states at distinct spatial locations. This spatially multiplexed polarization transformation approach was successfully exemplified through various designs, which showcased different applications, including polarization imaging and measurement[24], polarization field generation and vectorial holography[25–38], correction of polarization errors and aberrations[39], generation and conversion of beams with special polarization structures[40,41], and many others[42–44].

Despite these major advances, the current approaches to implement spatially multiplexed polarization transformations still confront certain challenges. Firstly, a large proportion of the polarization transformation systems utilizing metasurfaces can only accommodate input optical fields with a limited number of polarization modes or distributions that are known or pre-determined in the design phase. Stated differently, the polarization transformations assigned to these designs can only function when the input fields are known, causing them to be inapplicable to unknown polarization inputs, making their designs not universal. To overcome this constraint, researchers also reported metasurface-based designs to perform polarization transformations, acting on arbitrarily polarized input fields[41,45,46]. However, these designs still grapple with another limitation that their input fields must exhibit spatially uniform polarization distributions, implying that the input polarization state is confined to a maximum spatial degree of freedom of $N_i = 1$, even though their output polarization states are spatially varying with $N_o > 1$, as illustrated in **Fig. 1a**. Consequently, these designs can implement, at most, $1 \times N_o = N_o$ different polarization transformations from their input to output, and cannot process input fields with arbitrary spatially varying polarization states, i.e., cannot process distinct polarization states at different spatial locations ($N_i > 1$) at the input. In addition to these earlier efforts, there are also methods that utilized SLMs to manipulate input polarization fields with $N_i > 1$[38,39,47–51]. Nevertheless, for each of the polarization states within the input FOV, these SLM-based designs could only implement a single polarization transformation that maps to the same spatial location at the output FOV, and therefore these approaches could perform, at most, $N_i = N_o$ independent polarization transformations in a spatially parallel manner, as depicted in **Fig. 1b**.

To the best of our knowledge, within the existing body of literature concerning polarization transformations of arbitrary input polarization states, there is a gap wherein both the input and output polarization states can have spatial degrees of freedom that are cross-coupling with each other, simultaneously covering $N_i N_o$ polarization transformations in space, as depicted in **Fig. 1c**. In this case, the dense and complete mapping between the $N_i$ distinct input polarization states at the input FOV and the $N_o$ distinct output polarization states at the output FOV would permit $N_i N_o \gg 1$ unique polarization scattering matrices to be performed at the same time (i.e., through the same optical device); this would represent a universal polarization transformer, forming a much larger superset to all the earlier approaches reported to date, corresponding to **Figs. 1a and 1b**.

Here, we report a design strategy that uses a deep learning-optimized diffractive volume to perform universal polarization transformations between spatially varying polarization fields, all-optically implementing a large set of $N_i N_o$ complex-valued polarization scattering matrices spatially encoded within a compact space (**Fig. 1c)**. This diffractive volume is composed of multiple isotropic, spatially-engineered diffractive surfaces (layers) that contain thousands of diffractive features with trainable transmission coefficients, along with non-trainable, pre-determined arrays of linear polarizers (at 0°, 45°, 90° and 135°) positioned between these diffractive layers; see **Fig. 2**. Previously, the architecture of cascading such diffractive layers was demonstrated as an optical neural network, which can be used to all-optically perform machine learning inference and computing tasks, including, e.g., multiplexed linear transformations[52–71], and were also used for the inverse design of deterministic optical elements[72–74]. In this work, we used trainable diffractive layers and the fixed polarizer arrays as a compact diffractive polarization transformer, and programmed it to all-optically implement $N_i N_o$ distinct polarization scattering matrices that operate in parallel between different combinations of input and output pixels. Our numerical analyses revealed that when the number ($N$) of trainable diffractive features reaches $\geq 4 N_i N_o$, this diffractive universal polarization transformer can be trained to successfully approximate $N_i N_o$ spatially-encoded, target polarization scattering matrices between the input and output FOVs within a compact 3D volume. For example, we numerically showed that, within a short optical length of ~90×λ, this universal polarization transformer can all-optically and simultaneously synthesize $N_i N_o$=10,000 polarization scattering matrices, which can be further improved by increasing $N$.

Moreover, by integrating 3D-printed diffractive layers and wire-grid linear polarizers fabricated through photolithography, we successfully developed a proof-of-concept diffractive polarization transformer (operating at the terahertz part of the spectrum) that can all-optically perform polarization permutation operation of spatially varying input polarization fields with $N_i N_o$=16. Our experimental results revealed a good agreement with their numerically simulated counterparts, effectively showcasing the physical implementation of spatially-encoded independent polarization scattering matrices within a compact volume that axially spans 200×λ.

It is necessary to emphasize that the objective of our reported designs is *not* to generate a limited number of polarized optical fields from certain input fields, as such a goal is rather trivial and limited in scope. Instead, our diffractive processors are designed to synthesize **infinitely many** different output fields from **infinitely many** different input fields that possess arbitrary spatial distributions of polarization states, all following a given target set of $N_i N_o$ spatially-encoded polarization scattering matrices – forming a **universal polarization transformer** that is valid/accurate for infinitely many spatially varying polarization input fields. Furthermore, since these $N_i N_o$ polarization scattering matrices provide a *complete* description of the polarization transformations from each one of the polarization states located at different positions within the input FOV to each one of the points within the output FOV, they can be considered as the $N_i N_o$ distinct bases of all the possible polarization transformations between the spatially varying polarization fields within the input and output FOVs. **Therefore, the successful approximation of these $N_i N_o$ polarization scattering matrices marks the implementation of universal polarization transformations between an input FOV and an output FOV, which was never achieved or claimed by any design before this work**.

Our demonstrated framework also presents several additional advantages: (1) it relies on isotropic diffractive layers, which allows the use of standard dielectric materials with structured thickness profiles; (2) only a small number of polarization-encoded elements, e.g., linear polarizer arrays, are required in this design, which are readily available and can be easily integrated into existing imaging systems operating at different parts of the spectrum; (3) the desired polarization scattering matrices are all-optically synthesized based on solely passive materials through light-matter interaction, circumventing the requirements for preprocessing of information, digitization, and computing power; and (4) the designed framework is spatially scalable, making it adaptable to polarized optical fields with arbitrary apertures, operating at different parts of the electromagnetic spectrum.

With its unique capability of synthesizing an unprecedently large set of polarization scattering matrices (e.g., 10,000), this universal diffractive polarization transformer and the underlying design strategy will open up a rich space for complete control of the polarization-dependent impulse response function in optical imaging and sensing systems. Furthermore, the presented framework can be used to guide the design of novel optical devices that can perform the synthesis, modulation and characterization of polarization, as well as the development of intelligent machine vision systems with polarization-aware detection and classification capabilities. These advancements might inspire applications across numerous fields, including medical imaging, manufacturing, remote sensing and autonomous navigation.

## II. RESULTS AND DISCUSSION

### Formulism of universal polarization transformations

Throughout this manuscript, the terms "diffractive universal polarization transformer", "diffractive polarization transformer", and "diffractive processor" are interchangeably used. **Figure 1d** illustrates the general concept of a diffractive volume, with coherent illumination light assumed to propagate from the input plane to the output plane, where the input and output FOVs are defined with certain boundaries, covering the optical fields containing $N_i$ and $N_o$ distinct polarization states, respectively. We use the notation $\boldsymbol{i}$ and $\boldsymbol{o}$, respectively, to represent the sets of these input and output polarization states as a function of space. Based on this, we can write the $m^{\text{th}}$ and $n^{\text{th}}$ polarization states at the input and output polarization fields as:

$$\boldsymbol{i}^{(m)} = \begin{bmatrix} i_{\mathrm{x}}^{(m)} \\ i_{\mathrm{y}}^{(m)} \end{bmatrix}, \boldsymbol{o}^{(n)} = \begin{bmatrix} o_{\mathrm{x}}^{(n)} \\ o_{\mathrm{y}}^{(n)} \end{bmatrix} \tag{1}$$

where $i_{\mathrm{x}}^{(m)}$ and $i_{\mathrm{y}}^{(m)}$ represent the polarization components of $\boldsymbol{i}^{(m)}$ in the x and y directions, respectively, while $o_{\mathrm{x}}^{(n)}$ and $o_{\mathrm{y}}^{(n)}$ similarly represent the polarization components of $\boldsymbol{o}^{(n)}$ along the same directions ($1 \leq m \leq N_i$, $1 \leq n \leq N_o$). Without loss of generality, the representation of $\boldsymbol{i}^{(m)}$ and $\boldsymbol{o}^{(n)}$ can be converted to any other orthogonal polarization bases, such as left- or right-handed circular polarization, and this selection of polarization bases used in our analyses does not change any of our conclusions. When $N_i > 1$ and $N_o > 1$, each polarization state at the output FOV can be regarded as a complex-weighted linear combination (superposition) of all the polarization states distributed across the input FOV, which can be mathematically formulated as:

$$\boldsymbol{o}^{(n)} = \begin{bmatrix} o_{\mathrm{x}}^{(n)} \\ o_{\mathrm{y}}^{(n)} \end{bmatrix} = \sum_{m=1}^{N_i} \begin{bmatrix} S_{\mathrm{xx}}^{(m\to n)} & S_{\mathrm{yx}}^{(m\to n)} \\ S_{\mathrm{xy}}^{(m\to n)} & S_{\mathrm{yy}}^{(m\to n)} \end{bmatrix} \begin{bmatrix} i_{\mathrm{x}}^{(m)} \\ i_{\mathrm{y}}^{(m)} \end{bmatrix} = \sum_{m=1}^{N_i} \boldsymbol{S}^{(m\to n)} \boldsymbol{i}^{(m)} \tag{2}$$

where $\boldsymbol{S}^{(m\rightarrow n)} = \begin{bmatrix} S_{\mathrm{xx}}^{(m\rightarrow n)} & S_{\mathrm{yx}}^{(m\rightarrow n)} \\ S_{\mathrm{xy}}^{(m\rightarrow n)} & S_{\mathrm{yy}}^{(m\rightarrow n)} \end{bmatrix}$ is a 2×2 complex-valued polarization scattering matrix, which describes the complex-valued contributions from the $m^{\mathrm{th}}$ input polarization state $\boldsymbol{i}^{(m)}$ to the $n^{\mathrm{th}}$ output polarization state $\boldsymbol{o}^{(n)}$ through the diffractive volume. By gathering all the $\boldsymbol{S}^{(m\rightarrow n)}$ matrices that correspond to all the different combinations of the input and output polarization states, we can form a set $\mathbb{S} = \{\boldsymbol{S}^{(1\rightarrow 1)}, \dots, \boldsymbol{S}^{(m\rightarrow n)}, \dots, \boldsymbol{S}^{(N_i\rightarrow N_o)}\}$ that contains $N_i N_o$ spatially-encoded polarization scattering matrices, which constitute the bases for all the possible polarization transformations that can be performed by the diffractive volume, as illustrated in **Fig. 1e**.

Next, we formulate a mathematical representation of the transformation relationship between the input and output polarization fields within the diffractive volume. One way to formulate this relationship is to directly concatenate all the $\boldsymbol{o}^{(n)}$ vectors ($n \in [1, N_o]$) together to form a long vector $\overline{\boldsymbol{o}}$, which can be written as:

$$\overline{\boldsymbol{o}} = \begin{bmatrix} \boldsymbol{o}^{(1)} \\ \vdots \\ \boldsymbol{o}^{(N_o)} \end{bmatrix} = \begin{bmatrix} \boldsymbol{S}^{(1\rightarrow 1)} & \cdots & \boldsymbol{S}^{(N_i\rightarrow 1)} \\ \vdots & \ddots & \vdots \\ \boldsymbol{S}^{(1\rightarrow N_o)} & \cdots & \boldsymbol{S}^{(N_i\rightarrow N_o)} \end{bmatrix} \begin{bmatrix} \boldsymbol{i}^{(1)} \\ \vdots \\ \boldsymbol{i}^{(N_i)} \end{bmatrix} = \overline{\boldsymbol{S}}\,\overline{\boldsymbol{i}} \tag{3}$$

where $\overline{\boldsymbol{i}} = \begin{bmatrix} \boldsymbol{i}^{(1)} \\ \vdots \\ \boldsymbol{i}^{(N_i)} \end{bmatrix}$ and $\overline{\boldsymbol{o}} = \begin{bmatrix} \boldsymbol{o}^{(1)} \\ \vdots \\ \boldsymbol{o}^{(N_o)} \end{bmatrix}$ are complex-valued vectors describing the polarization fields within the input and output FOVs, respectively, which consist of $2N_i$ and $2N_o$ scalar elements that describe the $N_i$ and $N_o$ polarization states contained by the input and output FOVs, respectively. $\overline{\boldsymbol{S}} = \begin{bmatrix} \boldsymbol{S}^{(1\rightarrow 1)} & \cdots & \boldsymbol{S}^{(N_i\rightarrow 1)} \\ \vdots & \ddots & \vdots \\ \boldsymbol{S}^{(1\rightarrow N_o)} & \cdots & \boldsymbol{S}^{(N_i\rightarrow N_o)} \end{bmatrix}$ is a complex-valued matrix with a size of $2N_o \times 2N_i$, which includes all the $N_i N_o$ spatially-encoded polarization scattering matrices $\boldsymbol{S}^{(m\rightarrow n)}$ for all $m \in \{1, \dots, N_i\}$ and all $n \in \{1, \dots, N_o\}$.

Following this representation of polarization transformations, the entire $\overline{\boldsymbol{S}}$ matrix can be viewed as being composed of $N_i N_o$ blocks, each corresponding to a unique polarization scattering matrix. We can further separate the elements of $\overline{\boldsymbol{S}}$ based on their orthogonal polarization bases, achieved by formulating the output polarization field $\boldsymbol{o}$ with its x and y polarization components separated into two distinct vectors, denoted as $\boldsymbol{o}_{\mathrm{x}}$ and $\boldsymbol{o}_{\mathrm{y}}$, respectively, i.e.:

$$\boldsymbol{o}_{\mathrm{x}} = \begin{bmatrix} o_{\mathrm{x}}^{(1)} \\ \vdots \\ o_{\mathrm{x}}^{(N_o)} \end{bmatrix} = \begin{bmatrix} S_{\mathrm{xx}}^{(1\rightarrow 1)} & \cdots & S_{\mathrm{xx}}^{(N_i\rightarrow 1)} \\ \vdots & \ddots & \vdots \\ S_{\mathrm{xx}}^{(1\rightarrow N_o)} & \cdots & S_{\mathrm{xx}}^{(N_i\rightarrow N_o)} \end{bmatrix} \begin{bmatrix} i_{\mathrm{x}}^{(1)} \\ \vdots \\ i_{\mathrm{x}}^{(N_i)} \end{bmatrix} + \begin{bmatrix} S_{\mathrm{yx}}^{(1\rightarrow 1)} & \cdots & S_{\mathrm{yx}}^{(N_i\rightarrow 1)} \\ \vdots & \ddots & \vdots \\ S_{\mathrm{yx}}^{(1\rightarrow N_o)} & \cdots & S_{\mathrm{yx}}^{(N_i\rightarrow N_o)} \end{bmatrix} \begin{bmatrix} i_{\mathrm{y}}^{(1)} \\ \vdots \\ i_{\mathrm{y}}^{(N_i)} \end{bmatrix} \tag{4}$$

$$\boldsymbol{o}_{\mathrm{y}} = \begin{bmatrix} o_{\mathrm{y}}^{(1)} \\ \vdots \\ o_{y}^{(N_o)} \end{bmatrix} = \begin{bmatrix} S_{\mathrm{xy}}^{(1\to1)} & \cdots & S_{\mathrm{xy}}^{(N_i\to1)} \\ \vdots & \ddots & \vdots \\ S_{\mathrm{xy}}^{(1\to N_o)} & \cdots & S_{\mathrm{xy}}^{(N_i\to N_o)} \end{bmatrix} \begin{bmatrix} i_{\mathrm{x}}^{(1)} \\ \vdots \\ i_{\mathrm{x}}^{(N_i)} \end{bmatrix} + \begin{bmatrix} S_{\mathrm{yy}}^{(1\to1)} & \cdots & S_{\mathrm{yy}}^{(N_i\to1)} \\ \vdots & \ddots & \vdots \\ S_{\mathrm{yy}}^{(1\to N_o)} & \cdots & S_{\mathrm{yy}}^{(N_i\to N_o)} \end{bmatrix} \begin{bmatrix} i_{\mathrm{y}}^{(1)} \\ \vdots \\ i_{\mathrm{y}}^{(N_i)} \end{bmatrix} \quad (5)$$

Equations (4) and (5) can be further simplified by defining $\tilde{\boldsymbol{S}}_{\mathrm{xx}} = \begin{bmatrix} S_{\mathrm{xx}}^{(1\to1)} & \cdots & S_{\mathrm{xx}}^{(N_i\to1)} \\ \vdots & \ddots & \vdots \\ S_{\mathrm{xx}}^{(1\to N_o)} & \cdots & S_{\mathrm{xx}}^{(N_i\to N_o)} \end{bmatrix}$, $\tilde{\boldsymbol{S}}_{\mathrm{yx}} = \begin{bmatrix} S_{\mathrm{yx}}^{(1\to1)} & \cdots & S_{\mathrm{yx}}^{(N_i\to1)} \\ \vdots & \ddots & \vdots \\ S_{\mathrm{yx}}^{(1\to N_o)} & \cdots & S_{\mathrm{yx}}^{(N_i\to N_o)} \end{bmatrix}$, $\tilde{\boldsymbol{S}}_{\mathrm{xy}} = \begin{bmatrix} S_{\mathrm{xy}}^{(1\to1)} & \cdots & S_{\mathrm{xy}}^{(N_i\to1)} \\ \vdots & \ddots & \vdots \\ S_{\mathrm{xy}}^{(1\to N_o)} & \cdots & S_{\mathrm{xy}}^{(N_i\to N_o)} \end{bmatrix}$ and $\tilde{\boldsymbol{S}}_{\mathrm{yy}} = \begin{bmatrix} S_{\mathrm{yy}}^{(1\to1)} & \cdots & S_{\mathrm{yy}}^{(N_i\to1)} \\ \vdots & \ddots & \vdots \\ S_{\mathrm{yy}}^{(1\to N_o)} & \cdots & S_{\mathrm{yy}}^{(N_i\to N_o)} \end{bmatrix}$, which results in:

$$\boldsymbol{o}_{\mathrm{x}} = \tilde{\boldsymbol{S}}_{\mathrm{xx}}\boldsymbol{i}_{\mathrm{x}} + \tilde{\boldsymbol{S}}_{\mathrm{yx}}\boldsymbol{i}_{\mathrm{y}} \quad (6)$$

$$\boldsymbol{o}_{\mathrm{y}} = \tilde{\boldsymbol{S}}_{\mathrm{xy}}\boldsymbol{i}_{\mathrm{x}} + \tilde{\boldsymbol{S}}_{\mathrm{yy}}\boldsymbol{i}_{\mathrm{y}} \quad (7)$$

where $\tilde{\boldsymbol{S}}_{\mathrm{xx}}$, $\tilde{\boldsymbol{S}}_{\mathrm{yx}}$, $\tilde{\boldsymbol{S}}_{\mathrm{xy}}$ and $\tilde{\boldsymbol{S}}_{\mathrm{yy}}$ are all complex-valued matrices, each with a dimension of $N_o \times N_i$. $\boldsymbol{i}_{\mathrm{x}} = \begin{bmatrix} i_{\mathrm{x}}^{(1)} \\ \vdots \\ i_{\mathrm{x}}^{(N_i)} \end{bmatrix}$ and $\boldsymbol{i}_{\mathrm{y}} = \begin{bmatrix} i_{\mathrm{y}}^{(1)} \\ \vdots \\ i_{\mathrm{y}}^{(N_i)} \end{bmatrix}$ are both complex-valued vectors composed of $N_i$ elements that represent the x- and y-polarization components of the input polarization field $\boldsymbol{i}$ within the input FOV, respectively. Based on these definitions, we can write:

$$\boldsymbol{o} = \begin{bmatrix} \boldsymbol{o}_{\mathrm{x}} \\ \boldsymbol{o}_{\mathrm{y}} \end{bmatrix} = \begin{bmatrix} \tilde{\boldsymbol{S}}_{\mathrm{xx}} & \tilde{\boldsymbol{S}}_{\mathrm{yx}} \\ \tilde{\boldsymbol{S}}_{\mathrm{xy}} & \tilde{\boldsymbol{S}}_{\mathrm{yy}} \end{bmatrix} \begin{bmatrix} \boldsymbol{i}_{\mathrm{x}} \\ \boldsymbol{i}_{\mathrm{y}} \end{bmatrix} = \tilde{\boldsymbol{S}}\boldsymbol{i} \quad (8)$$

where $\boldsymbol{o} = \begin{bmatrix} \boldsymbol{o}_{\mathrm{x}} \\ \boldsymbol{o}_{\mathrm{y}} \end{bmatrix}$ is a complex-valued vector with $2N_o$ elements, which contains all the amplitude, phase and polarization information of the polarized output field within the output FOV, while $\boldsymbol{i} = \begin{bmatrix} \boldsymbol{i}_{\mathrm{x}} \\ \boldsymbol{i}_{\mathrm{y}} \end{bmatrix}$ is similar to $\boldsymbol{o}$ but for the input field, and is composed of $2N_i$ elements. $\tilde{\boldsymbol{S}} = \begin{bmatrix} \tilde{\boldsymbol{S}}_{\mathrm{xx}} & \tilde{\boldsymbol{S}}_{\mathrm{yx}} \\ \tilde{\boldsymbol{S}}_{\mathrm{xy}} & \tilde{\boldsymbol{S}}_{\mathrm{yy}} \end{bmatrix}$ represents a complex-valued matrix with a dimension of $2N_o \times 2N_i$, which contains all the $N_i N_o$ spatially-encoded polarization scattering matrices $\boldsymbol{S}^{(m\to n)}$ in the set $\mathbb{S}$, where the four entries in these $\boldsymbol{S}^{(m\to n)}$ matrices are separated into the four constituent submatrices of $\tilde{\boldsymbol{S}}$, i.e., $\tilde{\boldsymbol{S}}_{\mathrm{xx}}$, $\tilde{\boldsymbol{S}}_{\mathrm{yx}}$, $\tilde{\boldsymbol{S}}_{\mathrm{xy}}$ and $\tilde{\boldsymbol{S}}_{\mathrm{yy}}$, sharing the same relative locations within the submatrices.

**Design and numerical analyses of diffractive universal polarization transformers**

Next, we performed numerical analyses to demonstrate that, using a diffractive universal polarization transformer design shown in **Fig. 2**, a large number of polarization scattering matrices (e.g., 10,000) can be simultaneously implemented within a single diffractive volume. As depicted in **Fig. 2**, this diffractive polarization transformer comprises eight isotropic diffractive surfaces and two arrays of linear polarizers, expanding a total axial length of ~89.6λ. The linear polarizer arrays are positioned after the 3rd and 5th diffractive layers, so that the polarization modulation resulting from the polarizer arrays does not directly control the output polarization fields. Each of the polarizer arrays is pre-determined, composed of multiple linear polarizer units, with their polarization orientations at 0°, 45°, 90°, and 135°. These linear polarizer arrays act in synergy with the trainable isotropic diffractive layers to enable the diffractive polarization transformer to perform polarization-dependent modulation of the propagating complex fields. During the training process, the polarizer arrays are treated as non-trainable/fixed elements, while the transmission coefficients of the diffractive features on these diffractive surfaces constitute the only trainable parameters that can be updated through the error backpropagation. More details about the architecture, optical forward model, and the training hyperparameters of our diffractive polarization transformer designs are provided in the Methods Section.

In our numerical analyses, we selected $N_i = N_o = 10^2$, i.e., $\boldsymbol{i}_\mathrm{x}, \boldsymbol{i}_\mathrm{y}, \boldsymbol{o}_\mathrm{x}, \boldsymbol{o}_\mathrm{y} \in \mathbb{C}^{100\times1}$, and $\boldsymbol{i}, \boldsymbol{o} \in \mathbb{C}^{200\times1}$. Consequently, $\{\tilde{\boldsymbol{S}}_\mathrm{xx}, \tilde{\boldsymbol{S}}_\mathrm{yx}, \tilde{\boldsymbol{S}}_\mathrm{xy}, \tilde{\boldsymbol{S}}_\mathrm{yy}\} \in \mathbb{C}^{100\times100}$, and $\tilde{\boldsymbol{S}} \in \mathbb{C}^{200\times200}$. We randomly generated a complex-valued matrix with a size of 200 × 200 to serve as the ground truth for $\tilde{\boldsymbol{S}}$, which contains all the $N_i N_o =$ 10,000 spatially-encoded polarization scattering matrices $\boldsymbol{S}^{(m\to n)}$. The mutual differences between different submatrices $\{\tilde{\boldsymbol{S}}_\mathrm{xx}, \tilde{\boldsymbol{S}}_\mathrm{yx}, \tilde{\boldsymbol{S}}_\mathrm{xy}, \tilde{\boldsymbol{S}}_\mathrm{yy}\}$ are provided in **Supplementary Fig. S1**, confirming that the transformations between different polarization bases are all unique. The visualization of $\tilde{\boldsymbol{S}}$ in terms of the amplitude and phase components of the four constituent submatrices is also provided in **Fig. 3**. Next, we randomly generated 55,000 complex-valued vectors $\{\boldsymbol{i}\}$ as the input polarization fields with $N_i = 10^2$, and correspondingly constructed the set of output polarization fields $\{\boldsymbol{o}\}$ by calculating $\boldsymbol{o} = \tilde{\boldsymbol{S}}\boldsymbol{i}$ (see the Methods Section for details). We also randomly generated another set of 10,000 pairs of input-output fields as the blind testing set, not overlapping with the training fields. It is important to note that, due to the randomness in the generation process of $\tilde{\boldsymbol{S}}$ and $\{\boldsymbol{i}\}$, the resulting input and target output polarization fields in these training and testing sets are expected to exhibit a variety of elliptical polarizations, with a negligible probability of encountering linear and circular polarization states since the formation of these specific polarization states requires strict and specific constraints to be satisfied by the electric field components at the two polarization bases. Based on the given inputs fields $\{\boldsymbol{i}\}$, the objective of the training of our diffractive polarization transformer is to make its resulting output polarization fields $\{\boldsymbol{o}'\}$ come as close to the ground truth (target) output polarization field $\{\boldsymbol{o}\}$ as possible. Once this objective is achieved, the diffractive all-optical polarization transformations $\tilde{\boldsymbol{S}}'$ represented by the trained diffractive design can constitute an accurate approximation of the target polarization transformations $\tilde{\boldsymbol{S}}$, i.e., $\tilde{\boldsymbol{S}} \approx \tilde{\boldsymbol{S}}'$.

Following the configurations outlined above, we used deep learning to train various diffractive polarization transformer models using different numbers of trainable diffractive features, i.e., $N \in \{10.3\text{k} \approx N_i N_o; 20.0\text{k} \approx 2N_i N_o; 39.2\text{k} \approx 4N_i N_o; 80.0\text{k} \approx 8N_i N_o; 161\text{k} \approx 16N_i N_o; 320\text{k} \approx 32N_i N_o\}$, all using the same training dataset $\{(\boldsymbol{i}, \boldsymbol{o})\}$ and the same number of epochs. The loss function used for training these diffractive models is customized based on the mean-squared error (MSE) between the diffractive output fields $\{\boldsymbol{o}'\}$ and the ground truth (target) output fields $\{\boldsymbol{o}\}$ (see the Supplementary Information for details). After the convergence of the training, we measured the diffractive all-optical polarization transformations $\tilde{\boldsymbol{S}}'$ of these trained diffractive models, and quantified their transformation performance using four different metrics: (1) the normalized transformation MSE, denoted as $MSE_{\text{transform}}$; (2) the cosine similarity ($CosSim$) between the normalized versions of $\{\tilde{\boldsymbol{S}}'_{\text{xx}}, \tilde{\boldsymbol{S}}'_{\text{yx}}, \tilde{\boldsymbol{S}}'_{\text{xy}}, \tilde{\boldsymbol{S}}'_{\text{yy}}\}$, i.e., $\{\widehat{\boldsymbol{S}}'_{\text{xx}}, \widehat{\boldsymbol{S}}'_{\text{yx}}, \widehat{\boldsymbol{S}}'_{\text{xy}}, \widehat{\boldsymbol{S}}'_{\text{yy}}\}$, and the target transforms $\{\tilde{\boldsymbol{S}}_{\text{xx}}, \tilde{\boldsymbol{S}}_{\text{yx}}, \tilde{\boldsymbol{S}}_{\text{xy}}, \tilde{\boldsymbol{S}}_{\text{yy}}\}$; (3) MSE between the diffractive output fields and their ground truth fields, denoted as $MSE_{\text{output}}$; and (4) the minimum value of the polarization extinction ratio ($PER$) across the output FOV, denoted as $PER_{\text{min}}$, where $PER$ stands for the ratio between the optical field power at the desired polarization state and the undesired, orthogonal polarization state. More details about the implementation of the training and these performance metrics can be found in the Methods Section.

Using these four evaluation metrics, we quantified the performance of our trained diffractive polarization transformer models, and reported the results in **Fig. 4** as average values across the entire testing set as a function of the number of diffractive features *N* in each design. As illustrated in **Figs. 4a and b**, when *N* approaches a threshold of 39.2k ≈ $4N_i N_o$, the transformation errors for all the submatrices converge towards 0, while the cosine similarity simultaneously approaches 1. These results indicate that the all-optically implemented transformation matrices $\{\widehat{\boldsymbol{S}}'_{\text{xx}}, \widehat{\boldsymbol{S}}'_{\text{yx}}, \widehat{\boldsymbol{S}}'_{\text{xy}}, \widehat{\boldsymbol{S}}'_{\text{yy}}\}$ successfully approximate their respective target polarization transformations $\{\tilde{\boldsymbol{S}}_{\text{xx}}, \tilde{\boldsymbol{S}}_{\text{yx}}, \tilde{\boldsymbol{S}}_{\text{xy}}, \tilde{\boldsymbol{S}}_{\text{yy}}\}$ with negligible error, provided that $N \geq 4N_i N_o$. This finding aligns with the fact that all the four submatrices of $\tilde{\boldsymbol{S}}$ (i.e., $\tilde{\boldsymbol{S}}_{\text{xx}}$, $\tilde{\boldsymbol{S}}_{\text{yx}}$, $\tilde{\boldsymbol{S}}_{\text{xy}}$ and $\tilde{\boldsymbol{S}}_{\text{yy}}$) can be arbitrarily selected, with a total of $4N_i N_o$ independent variables, which in turn mandates that the diffractive model must have sufficient degrees of freedom ($\geq 4N_i N_o$) to completely cover the variable space of the target polarization transformations. Moreover, our success is also reflected in the accurate approximation of all the 10,000 target polarization scattering matrices, $\mathbb{S} = \{\boldsymbol{S}^{(1\to1)}, \dots, \boldsymbol{S}^{(m\to n)}, \dots, \boldsymbol{S}^{(N_i\to N_o)}\}$. As demonstrated in **Supplementary Fig. S2**, the cosine similarity values between the scattering matrices $\boldsymbol{S}'^{(1\to1)}, \dots, \boldsymbol{S}'^{(m\to n)}, \dots, \boldsymbol{S}'^{(N_i\to N_o)}$ all-optically implemented using our diffractive polarization transformer with $N = 4N_iN_o$ and their corresponding ground truth scattering matrices $\boldsymbol{S}^{(1\to1)}, \ldots, \boldsymbol{S}^{(m\to n)}, \ldots, \boldsymbol{S}^{(N_i\to N_o)}$ were found to be ~1 (all greater than 0.9999), demonstrating successful approximation of all the $N_iN_o$ = 10,000 target polarization scattering matrices through the same diffractive processor. Moreover, in **Fig. 4e**, $MSE_{\text{output}}$ exhibits a similar trend as the performance metrics presented in **Fig. 4a**, indicating a monotonous decrease to 0 as $N$ approaches $4N_iN_o$. As illustrated in **Fig. 4f**, $PER_{\min}$ also confirms the same trends, and shows an increase as $N$ increases, with a particularly significant jump in performance when $N \approx 4N_iN_o$. These findings underscore the significance of $N = 4N_iN_o$ as the turning point for a diffractive polarization transformer to accurately approximate all the $N_iN_o$ = 10,000 target polarization scattering matrices with negligible error.

It is worth noting that, **Figs. 4a and b** reveal some imbalance among the transformation approximation accuracies corresponding to the four submatrices $\{\tilde{\boldsymbol{S}}_{\text{xx}}, \tilde{\boldsymbol{S}}_{\text{yx}}, \tilde{\boldsymbol{S}}_{\text{xy}}, \tilde{\boldsymbol{S}}_{\text{yy}}\}$ when $N < 4N_iN_o$; for example, the approximation error corresponding to $\tilde{\boldsymbol{S}}_{\text{yy}}$ is the smallest among the four. This performance imbalance observed under $N < 4N_iN_o$ is actually caused by the geometric asymmetry within the polarizer arrays used in our diffractive processor design. This asymmetric behavior, however, does not cause an output performance discrepancy between the x and y polarization states of the output field. To shed more light on this, as shown in **Figs. 4c and d,** we averaged the performance metrics of the submatrices in **Figs. 4a and b** that correspond to the same polarization bases at the output (i.e., x and y), generating the averaged transformation MSE, $\frac{1}{2}[MSE_{\text{transform}}(\tilde{\boldsymbol{S}}_{\text{xx}},) + MSE_{\text{transform}}(\tilde{\boldsymbol{S}}_{\text{yx}},)]$ and $\frac{1}{2}[MSE_{\text{transform}}(\tilde{\boldsymbol{S}}_{\text{xy}}) + MSE_{\text{transform}}(\tilde{\boldsymbol{S}}_{\text{yy}})]$, as well as the averaged transformation cosine similarity values, $\frac{1}{2}[CosSim(\tilde{\boldsymbol{S}}_{\text{xx}}) + CosSim(\tilde{\boldsymbol{S}}_{\text{yx}})]$ and $\frac{1}{2}[CosSim(\tilde{\boldsymbol{S}}_{\text{xy}}) + CosSim(\tilde{\boldsymbol{S}}_{\text{yy}})]$. These average MSE and cosine similarity values are very close to each other for x and y polarization states (see **Figs. 4c-d**), indicating that our diffractive polarization transformer designs, even for $N < 4N_iN_o$, successfully managed to strike a balance in the performance of polarization transformations, without introducing an accuracy imbalance between the x and y polarization components at the output field.

Next, we performed further blind testing using input fields that exhibit only linear and circular polarization states, which were never seen by these diffractive models during the training process. For each polarization type (linear-only and circular-only), we randomly generated 10,000 different input fields and accordingly obtained their ground truth output fields by calculating $\boldsymbol{o} = \tilde{\boldsymbol{S}}\boldsymbol{i}$ (see the Methods Section for details). Using these newly generated input polarized fields, we conducted additional tests on our trained diffractive designs with $N = 8N_iN_o$, the results of which are summarized in **Fig. 5**. To provide a more intuitive and quantitative evaluation of the quality of the all-optically synthesized polarization states at the output fields, we focused on two quantities: (1) the amplitude ratio between the electric field components in the x and y directions, expressed as $\text{atan}\frac{|\boldsymbol{E}_\text{y}|}{|\boldsymbol{E}_\text{x}|}$, where $|\boldsymbol{E}_\text{x}|$ and $|\boldsymbol{E}_\text{y}|$ represent the amplitude of the x and y polarization components of the optical field $\boldsymbol{E}$ (with $\text{atan}\frac{|\boldsymbol{E}_\text{y}|}{|\boldsymbol{E}_\text{x}|} \in [0, \frac{\pi}{2}]$); and (2) the wrapped phase difference between $\boldsymbol{E}_\text{x}$ and $\boldsymbol{E}_\text{y}$, i.e., $|\angle\boldsymbol{E}_\text{y} - \angle\boldsymbol{E}_\text{x}|_\pi$. The results of our analysis (**Fig. 5**) using the diffractive polarization transformer design with $N = 8N_iN_o$ reveal that, regardless of the different polarization states utilized at the input, all the corresponding output polarization states exhibit a very good match with their ground truth counterparts with minimal error. Moreover, the *PER* values of all the pixels in the output FOV exceed 40 dB (**Fig. 5**, right panel), further confirming the capability of our diffractive polarization processor to effectively perform universal polarization transformations of spatially varying polarized optical fields.

**Experimental validation of a diffractive polarization transformer**

We performed a proof-of-concept experimental demonstration of our diffractive universal polarization transformation framework by designing a diffractive processor to perform a random permutation operation of polarization fields. Based on a continuous-wave terahertz (THz) set-up shown in **Fig. 6a** that uses an illumination wavelength of λ = 0.75 mm, we selected $N_i = N_o = 2^2$ (i.e., $\boldsymbol{i}_\text{x}, \boldsymbol{i}_\text{y}, \boldsymbol{o}_\text{x}, \boldsymbol{o}_\text{y} \in \mathbb{C}^{4\times1}$ and $\boldsymbol{i}, \boldsymbol{o} \in \mathbb{C}^{8\times1}$), and designed a diffractive polarization transformer composed of three isotropic, phase-only diffractive layers ($L_1$-$L_3$) and two polarizer arrays ($PA_1$ and $PA_2$), covering a total axial length of ~200λ. The target polarization transformations $\tilde{\boldsymbol{S}}$ are defined by a randomly generated 8×8 permutation matrix shown in **Fig. 7a, left**, i.e., $\tilde{\boldsymbol{S}} \in \mathbb{R}^{8\times8}$ and $\{\tilde{\boldsymbol{S}}_\text{xx}, \tilde{\boldsymbol{S}}_\text{yx}, \tilde{\boldsymbol{S}}_\text{xy}, \tilde{\boldsymbol{S}}_\text{yy}\} \in \mathbb{R}^{4\times4}$. A polarization permutation transformation treats the x and y components of each spatial pixel in the input polarization field separately, which are then permuted (spatially shuffled) with respect to both the spatial distribution and the polarization state before being projected onto the output plane of the diffractive processor. To implement these target polarization transformations ($\tilde{\boldsymbol{S}}$ matrix shown in **Fig. 7a, left**), we randomly generated a total of 40,000 input-output polarization field pairs ($\boldsymbol{i}$ and $\boldsymbol{o}$) that all satisfy $\boldsymbol{o} = \tilde{\boldsymbol{S}}\boldsymbol{i}$, and subsequently used them as the training data to optimize the thickness values of the three diffractive layers ($L_1$-$L_3$) during the training process. In order to mitigate a potential performance degradation that could result from misalignment errors due to, e.g., imperfect assembly of the diffractive layers, we also adopted a "vaccination" training strategy[56,57] where the random displacement of these diffractive layers was modeled and incorporated as random errors into the physical forward model used during the training process.

After the completion of the training phase, we utilized a 3D printer to fabricate the resulting diffractive layers, and the photographs of the fabricated components are presented in **Fig. 6d**. We also fabricated the two polarizer arrays utilizing a photolithography process, which entailed the coating of 8×8 THz wire-grid linear polarizers with varying polarizing angles onto a fused silica wafer. The design layout and the finished component after the fabrication of the polarizer array $PA_1$ are shown in **Fig. 6c**; the orientations of the linear polarizers in $PA_2$ are 180-degree rotated versions of those in $PA_1$. These polarizer arrays were then assembled with the fabricated diffractive layers using a custom-designed 3D-printed holder, ultimately forming our physical diffractive polarization transformer, as shown in the inset of **Fig. 6a**. More details regarding the physical architecture, training data generation, vaccinated training strategy and fabrication process of this experimental diffractive design are reported in the Methods Section and **Supplementary Fig. S6**.

To experimentally evaluate the effectiveness of our 3D-fabricated diffractive polarization transformer, we illuminated, one by one, the pixels in the input FOV using a linear polarization of x or y, and measured the output power of all the pixels in the output FOV at both x and y polarization states by rotating a THz wire-grid analyzer. By performing these measurements iteratively for all the spatial pixels within the input FOV using both polarization states (i.e., x and y), we were able to experimentally synthesize all the resulting output polarized fields, measuring the all-optical polarization transformations ($\tilde{\boldsymbol{S}}'$) represented by our trained diffractive design. The results of this analysis are summarized in **Fig. 7a, right**, which reveals that the experimentally measured diffractive all-optical polarization transformations $\tilde{\boldsymbol{S}}'^{(\text{exp})}$ exhibit minimal differences from their numerically simulated counterparts (**Fig. 7a, middle**), also demonstrating a good agreement with the ground truth (**Fig. 7a, left**), i.e., $\tilde{\boldsymbol{S}} \approx \tilde{\boldsymbol{S}}' \approx \tilde{\boldsymbol{S}}'^{(\text{exp})}$.

Furthermore, we conducted additional experiments to demonstrate the capability of our diffractive polarization transformer in performing the target polarization permutations of the input fields with more complex polarization distributions. To achieve this, we carried out three different types of operation at the input: (1) illuminating a single pixel as before, but using both x and y polarization states simultaneously; (2) illuminating two pixels at different spatial positions while using the same polarization (e.g., x polarization); and (3) illuminating two pixels at different spatial positions with distinct polarizations, i.e., one pixel with x polarization and the other pixel with y polarization. These three scenarios correspond to the input polarization fields shown in the 1st, 2nd, and 3rd to 5th rows of **Fig. 7b**, respectively, where their corresponding numerically simulated and experimentally measured output fields were also presented on the right side of the same rows, revealing a good match with each other. The success of these experimental results further confirms the feasibility of our diffractive polarization transformation framework.

**Discussion**

We demonstrated that a diffractive polarization transformer could be designed to perform a large set of spatially encoded polarization scattering matrices, performing universal polarization transformations between the input and output optical fields with spatially varying polarization. To the best of our knowledge, this is the first demonstration of a design to all optically implement universal polarization transformations between two optical fields with spatially varying polarization properties, which corresponds to the case depicted in **Fig. 1c**.

In addition to all-optically implementing an arbitrary $\tilde{\boldsymbol{S}}$ matrix using our diffractive processor design depicted in **Fig. 2**, we would like to emphasize that there are certain classes of linear transformations of input polarization fields that could be implemented without utilizing any polarization-sensitive elements, such as linear polarizer arrays. In this simplified isotropic diffractive polarization transformer without any polarizer arrays, the achievable set of linear transformations can be represented as $\tilde{\boldsymbol{S}} = \begin{bmatrix} \tilde{\boldsymbol{S}}_{\mathrm{LT}} & \boldsymbol{0} \\ \boldsymbol{0} & \tilde{\boldsymbol{S}}_{\mathrm{LT}} \end{bmatrix}$, where $\tilde{\boldsymbol{S}}_{\mathrm{LT}}$ is a complex-valued matrix that can be arbitrarily selected. This particular form of $\tilde{\boldsymbol{S}}$, as compared to the general form that we used previously, is characterized by the additional constraints of $\tilde{\boldsymbol{S}}_{\mathrm{xx}} = \tilde{\boldsymbol{S}}_{\mathrm{yy}} = \tilde{\boldsymbol{S}}_{\mathrm{LT}}$ and $\tilde{\boldsymbol{S}}_{\mathrm{xy}} = \tilde{\boldsymbol{S}}_{\mathrm{yx}} = \boldsymbol{0}$. These constraints arise from the lack of polarization-sensitive elements in this simplified diffractive processor design without any polarizer arrays, which limits the ability of the diffractive processor to perform unique manipulation of distinct polarization states, and leads to an isotropic linear transformation that exhibits identical responses to $\boldsymbol{i}_{\mathrm{x}}$ and $\boldsymbol{i}_{\mathrm{y}}$. This transformation can still be used for the processing of polarization fields, resulting in a linear combination of various polarization states that are spatially located at different positions within the input polarized field. To numerically demonstrate the efficacy of this specialized transformation, we randomly generated an $\tilde{\boldsymbol{S}}_{\mathrm{LT}} \in \mathbb{C}^{100\times100}$ to form the target transformation $\tilde{\boldsymbol{S}}$, and visualized the phase and amplitude components of $\tilde{\boldsymbol{S}}_{\mathrm{LT}}$ in **Supplementary Fig. S3**. Using the same architecture and deep learning-based training strategy of the diffractive processor designs shown in **Fig. 2** (but without employing the polarizer arrays), we trained three diffractive models with $N = 0.5N_iN_o$, $N = N_iN_o$, $N = 2N_iN_o$ to all-optically approximate the new target transformation $\tilde{\boldsymbol{S}}$. To evaluate the approximation performance of this diffractive processor, we employed the same metrics $MSE_{\mathrm{transform}}$, $CosSim$, $MSE_{\mathrm{output}}$ and $PER_{\mathrm{min}}$ used in our earlier analyses, and summarized the results of these performance metrics as a function of $N$; see **Supplementary Fig. S4**. We also showed the resulting $\widehat{\boldsymbol{S}}'_{\mathrm{LT}}$ of the diffractive processor designs in **Supplementary Fig. S3**, and visualized exemplary output polarization fields $\widehat{\boldsymbol{o}}'$ synthesized using the diffractive model with $N = N_iN_o$ that was tested based on the input fields generated with customized types of polarization states (i.e., linear-only, circular-only and elliptical); all of these results provide a remarkable similarity to their ground truth with negligible error (see **Supplementary Fig. S5**), demonstrating the successful all-optical implementation of the target spatial transformations of the polarization fields.

Finally, we would like to emphasize that our diffractive universal polarization transformation framework is inherently scalable. Based on a diffractive polarization transformer model trained using a dedicated operating wavelength, the design of these diffractive layers and polarizer arrays can be further scaled up or down (i.e., stretched or shrank) to process polarization fields at another illumination wavelength, without the need to retrain its design, making this framework applicable to operate at different parts of the electromagnetic spectrum, including the visible band. While the diffractive polarization transformers presented in this manuscript used monochromatic illumination, they can also be integrated with broadband processor designs that were reported earlier[60,72,73,71], which can be used to simultaneously process the amplitude, phase, polarization and spectral information of the input optical fields in unique ways that are not possible with earlier designs. We also believe that our reported framework can create polarization-dependent transfer functions for designing optical systems, which might find various applications in e.g., detection and classification of samples with unique polarization and chiral properties[7,8,75], compressive encoding and encryption of polarization information[34,76,77], and correction of polarization-related aberrations and errors[78–80]. Moreover, we envision that the capabilities of our diffractive processor to transform polarization states across a 2D transverse plane can be further extended to the transformation of the axial polarization states along the direction of the light propagation[41], which might open up new opportunities in, e.g., particle manipulation[81] and super-resolution imaging[82].

## III. METHODS SECTION

### Numerical forward model of the diffractive polarization transformer

The Jones vector of the electric field $\boldsymbol{E}$ at a spatial location $(x, y, z)$ in a fully polarized optical field propagating along the z-axis can be written using the following form:

$$\boldsymbol{E}(x,y,z) = \begin{bmatrix} E_{\mathrm{x}}(x,y,z) \\ E_{\mathrm{y}}(x,y,z) \end{bmatrix} \tag{9}$$

where $E_{\mathrm{x}}$ and $E_{\mathrm{y}}$ represent the scalar electric field components at x and y, respectively. Since the isotropic diffractive layers are not polarization-sensitive, their resulting complex-valued modulation is identical for both of the orthogonal polarization components. The complex-valued transmission coefficient of the $m^{\mathrm{th}}$ feature on the $k^{\mathrm{th}}$ diffractive layer at spatial location $(x_m, y_m, z_k)$ can be formulated as:

$$t_m(x_m, y_m, z_k) = a^k(x_m, y_m, z_k)\exp\left(j\phi^k(x_m, y_m, z_k)\right) \tag{10}$$

where $a^k$ and $\phi^k$ denote the amplitude and phase coefficients of the $m^{\mathrm{th}}$ diffractive feature, which are both trainable with a permitted range of $[0, 1]$ and $[0, 2\pi]$, respectively. The size of each diffractive feature on the transmissive surfaces is selected as ~$0.5\lambda$, and the optical field pixels at the input/output FOVs are spatially binned to have a size of ~$\lambda$.

Between the successive diffractive layers, the optical fields are assumed to propagate in air (n = 1), and the modulation of these fields is calculated by performing free-space propagation of each of the orthogonal polarization components using the Rayleigh-Sommerfeld diffraction:

$$w(x, y, z) = \frac{z}{r^2}\left(\frac{1}{2\pi r^2} + \frac{1}{j\lambda}\right)\exp\left(\frac{j2\pi r}{\lambda}\right) \tag{11}$$

where $r = \sqrt{x^2 + y^2 + z^2}$ and $j = \sqrt{-1}$. Therefore, the scalar complex field $E_k(x_m, y_m, z_m)$ right after the $k^{\text{th}}$ diffractive layer located at $z = z_k$ can be written as:

$$E_k(x_m, y_m, z_k) = t_m(x_m, y_m, z_k) \cdot \sum_{n \in \mathbb{N}} E_{k-1}(x_n, y_n, z_{k-1})\, w(x_m, y_m, \Delta z_k) \tag{12}$$

where $\mathbb{N}$ denotes all the diffractive features on the $(k-1)^{\text{th}}$ diffractive surface, and $\Delta z_k = z_k - z_{k-1}$ is the axial distance between two successive layers (including the diffractive surfaces and input/output planes). In our implementation, $\Delta z_k$ is empirically selected as $0.27 D_{\text{Layer}}$, where $D_{\text{Layer}}$ represents the lateral size/width of each diffractive layer and is determined based on the selection of $N$.

For the polarization transformers used in this work, the linear polarizer array is modeled using a spatial location-dependent Jones matrix $\boldsymbol{J}_{\text{linear}}(x, y, z)$ to simulate its modulation of the polarized electric field at the location $(x, y, z)$, which can be formulated as:

$$\boldsymbol{E}_{\text{out}}(x, y, z) = \boldsymbol{J}_{\text{linear}}(x, y, z)\boldsymbol{E}_{\text{in}}(x, y, z) \tag{13}$$

where $\boldsymbol{E}_{\text{in}}$ and $\boldsymbol{E}_{\text{out}}$ denote the incident and exiting polarized electric fields at the linear polarizer array, respectively. $\boldsymbol{J}_{\text{linear}}(x, y, z)$ is given by:

$$\boldsymbol{J}_{\text{linear}}(x, y, z) = \begin{bmatrix} \cos^2\theta(x, y, z) & \cos\theta(x, y, z)\sin\theta(x, y, z) \\ \cos\theta(x, y, z)\sin\theta(x, y, z) & \sin^2\theta(x, y, z) \end{bmatrix} \tag{14}$$

where $\theta(x, y, z)$ represents the angle between the x-axis and the polarization axis of the linear polarizer located at $(x, y, z)$ on the plane of the linear polarizer array. As illustrated in **Fig. 2**, the pre-determined polarizer arrays in our design contain in total four types of linear polarizer units with 4 different polarization directions, $\theta = \{0, 0.25\pi, 0.5\pi, 0.75\pi\}$. These 4 different types of linear polarizers are spatially binned to have a 2 × 2 period and tiled along the x and y directions, forming a repeating array. The side length of each linear polarizer array is designed to coarsely match that of the diffractive layers; for example, in the diffractive model with $N = 8N_iN_o$, the linear polarizer array has a size of ~$53.3\lambda$. The residual space surrounding the polarizer arrays is assumed to be fully transmissive without any polarization modulation. For all the diffractive designs used for numerical analyses in this manuscript, we empirically set the axial distance between a given polarizer array and the adjacent diffractive layer in front of them as 0, i.e., $d_{p1} = d_{p2} = 0$ in **Fig. 2**.

**Preparation of the polarization transformation datasets**

In our diffractive polarization transformer design shown in **Fig. 2**, the input and output FOVs are assumed to have the same size of 10 × 10 pixels, covering an area of ~$10\lambda \times 10\lambda$. The polarization components of the input and output complex fields within these input and output FOVs can be represented using flattened vectors with a size of 100 × 1, i.e., $\{\boldsymbol{i}_{\mathrm{x}}, \boldsymbol{i}_{\mathrm{y}}, \boldsymbol{o}_{\mathrm{x}}, \boldsymbol{o}_{\mathrm{y}}\} \in \mathbb{C}^{100\times1}$, hence the vectors $\boldsymbol{i} = \begin{bmatrix}\boldsymbol{i}_{\mathrm{x}}\\ \boldsymbol{i}_{\mathrm{y}}\end{bmatrix}$ and $\boldsymbol{o} = \begin{bmatrix}\boldsymbol{o}_{\mathrm{x}}\\ \boldsymbol{o}_{\mathrm{y}}\end{bmatrix}$ that represent the input and output polarized fields have a size of 200 × 1, i.e., $\{\boldsymbol{i} = \begin{bmatrix}\boldsymbol{i}_{\mathrm{x}}\\ \boldsymbol{i}_{\mathrm{y}}\end{bmatrix}, \boldsymbol{o} = \begin{bmatrix}\boldsymbol{o}_{\mathrm{x}}\\ \boldsymbol{o}_{\mathrm{y}}\end{bmatrix}\} \in \mathbb{C}^{200\times1}$. For creating the polarization transformation matrix $\tilde{\boldsymbol{S}}$, the amplitude and phase components of the matrix elements were all generated with a uniform distribution of $U[0, 1]$ and $U[0, 2\pi]$, respectively. Next, the amplitude and phase components of the input fields $\boldsymbol{i} = \begin{bmatrix}\boldsymbol{i}_{\mathrm{x}}\\ \boldsymbol{i}_{\mathrm{y}}\end{bmatrix}$ were also randomly generated with a uniform ($U$) distribution of $U[0,1]$ and $U[0, 2\pi]$, respectively. The ground truth (target) fields $\boldsymbol{o}$ were generated by calculating $\boldsymbol{o} = \begin{bmatrix}\boldsymbol{o}_{\mathrm{x}}\\ \boldsymbol{o}_{\mathrm{y}}\end{bmatrix} = \tilde{\boldsymbol{S}}\boldsymbol{i} = \tilde{\boldsymbol{S}}\begin{bmatrix}\boldsymbol{i}_{\mathrm{x}}\\ \boldsymbol{i}_{\mathrm{y}}\end{bmatrix}$. For each transformation task of interest, we generated a total of 70,000 input/output complex fields to form a dataset, divided into three parts: training, validation, and testing, each containing 55,000, 5,000 and 10,000 complex-valued field pairs, respectively. In addition, we also prepared data where the input fields have special polarization states. For example, we generated input fields $\boldsymbol{i}$ that are purely linearly polarized by adding a pixel-wise constraint of $\left|\angle\boldsymbol{i}_{\mathrm{x}} - \angle\boldsymbol{i}_{\mathrm{y}}\right|_{\pi} = 0$ or $\pi$ ($|\cdot|_{\pi}$ indicates the phase wrapping operation) during the data generation. We also generated circularly polarized input fields by adding pixel-wise constraints of $|\boldsymbol{i}_{\mathrm{x}}| = |\boldsymbol{i}_{\mathrm{y}}|$ and $\left|\angle\boldsymbol{i}_{\mathrm{x}} - \angle\boldsymbol{i}_{\mathrm{y}}\right|_{\pi} = \frac{\pi}{2}$ or $\frac{3\pi}{2}$. For each of these special polarization types, in total 10,000 complex-valued field pairs were generated for blind testing.

For our experimental design of the diffractive polarization transformer shown in **Fig. 6** and **Supplementary Fig. S6**, the input and output FOVs have the same size of 2 × 2 pixels. Therefore, the polarization components of the input and output complex fields within these FOVs can be represented using flattened vectors with a size of 4 × 1, i.e., $\{\boldsymbol{i}_{\mathrm{x}}, \boldsymbol{i}_{\mathrm{y}}, \boldsymbol{o}_{\mathrm{x}}, \boldsymbol{o}_{\mathrm{y}}\} \in \mathbb{C}^{4\times1}$, hence the vectors $\boldsymbol{i} = \begin{bmatrix}\boldsymbol{i}_{\mathrm{x}}\\ \boldsymbol{i}_{\mathrm{y}}\end{bmatrix}$ and $\boldsymbol{o} = \begin{bmatrix}\boldsymbol{o}_{\mathrm{x}}\\ \boldsymbol{o}_{\mathrm{y}}\end{bmatrix}$ that represent the input and output polarized fields have a size of 8 × 1, i.e., $\{\boldsymbol{i} = \begin{bmatrix}\boldsymbol{i}_{\mathrm{x}}\\ \boldsymbol{i}_{\mathrm{y}}\end{bmatrix}, \boldsymbol{o} = \begin{bmatrix}\boldsymbol{o}_{\mathrm{x}}\\ \boldsymbol{o}_{\mathrm{y}}\end{bmatrix}\} \in \mathbb{C}^{8\times1}$. The target matrix $\tilde{\boldsymbol{S}}$ was a randomly generated 8 × 8 permutation matrix, as shown in **Fig. 7a, left**. We generated a total of 55,000 input/output complex field pairs to form a dataset, and further divided it into three parts: training, validation, and testing, each containing 40,000, 5,000 and 10,000 complex-valued field pairs, respectively. Of the 40,000 training samples, the first 20,000 were created by randomly generating the amplitude and phase components of the input fields $\boldsymbol{i} = \begin{bmatrix} \boldsymbol{i}_\mathrm{x} \\ \boldsymbol{i}_\mathrm{y} \end{bmatrix}$ using a uniform ($U$) distribution of $U[0.2, 1]$ and $U[0, 2\pi]$, respectively, and then generating the target fields $\boldsymbol{o}$ through calculating $\boldsymbol{o} = \begin{bmatrix} \boldsymbol{o}_\mathrm{x} \\ \boldsymbol{o}_\mathrm{y} \end{bmatrix} = \tilde{\boldsymbol{S}}\boldsymbol{i} = \tilde{\boldsymbol{S}} \begin{bmatrix} \boldsymbol{i}_\mathrm{x} \\ \boldsymbol{i}_\mathrm{y} \end{bmatrix}$. The second 20,000 training samples were generated in a way identical to the first 20,000 samples, but with one exception: a random portion (up to 7) of the entries in the input fields $\boldsymbol{i}$ were deliberately set to 0. This structuring of the training data helped us enhance the blind testing performance of our experimental diffractive model when processing sparse input fields, which were used to characterize its polarization transformation performance.

**Performance metrics used for the quantification of universal polarization transformations**

To quantitatively evaluate the results of our diffractive polarization transformers, four different performance metrics were calculated using the blind testing data sets: (1) $MSE_\mathrm{transform}$, (2) $CosSim$ between an all-optical complex-valued polarization transformation matrix and its ground truth, (3) $MSE_\mathrm{output}$ between the diffractive output polarized fields and their ground truth, and (4) $PER_\mathrm{min}$. The transformation error $MSE_\mathrm{transform}$ is defined as:

$$MSE_\mathrm{transform} = \frac{1}{N_i N_o} \sum_{n=1}^{N_i N_o} |\tilde{\boldsymbol{s}}[n] - m\tilde{\boldsymbol{s}}'[n]|^2 = \frac{1}{N_i N_o} \sum_{n=1}^{N_i N_o} |\tilde{\boldsymbol{s}}[n] - \hat{\boldsymbol{s}}'[n]|^2 \quad (15)$$

where $\tilde{\boldsymbol{s}}$ represents the vectorized version of the ground truth (target) transformation matrix $\tilde{\boldsymbol{S}}$, and $\tilde{\boldsymbol{s}}'$ is the vectorized version of the diffractive all-optical transformation matrix $\tilde{\boldsymbol{S}}'$, with its normalized version denoted as $\hat{\boldsymbol{s}}'$. $m$ is a scalar normalization coefficient used to eliminate the effect of diffraction efficiency-related scaling mismatch between $\tilde{\boldsymbol{S}}$ and $\tilde{\boldsymbol{S}}'$,[83] i.e.,

$$m = \frac{\sum_{n=1}^{N_i N_o} \tilde{\boldsymbol{s}}[n]\boldsymbol{s}'^{*}[n]}{\sum_{n=1}^{N_i N_o} |\tilde{\boldsymbol{s}}'[n]|^2} \quad (16).$$

The cosine similarity between the all-optical polarization transform and the target transform is defined as:

$$CosSim = \frac{|\boldsymbol{s}^H \hat{\boldsymbol{s}}'\,|}{\sqrt{\sum_{n=1}^{N_i N_o} |\boldsymbol{s}[n]|^2}\sqrt{\sum_{n=1}^{N_i N_o} |\hat{\boldsymbol{s}}'[n]|^2}} \quad (17)$$

The polarization extinction ratio, $PER$, is defined as:

$$PER[n](dB) = 20\log_{10}\frac{|o_x^*[n]o_x'[n] + o_y^*[n]o_y'[n]|}{|o_y[n]o_x'[n] - o_x[n]o_y'[n]|} \quad (18)$$

Here, $o_\#'[n]$ represents the $n^{\text{th}}$ element of the diffractive output field at the polarization state of #, and $o_\#[n]$ stands for the $n^{\text{th}}$ element of the corresponding target (ground truth) output field at the polarization state of #.

The normalized mean-squared error $MSE_{\text{output}}$ between the diffractive outputs and their ground truth is defined using the same formula as the loss function used for the training process (see the Supplementary Information).

Finally, the diffraction efficiency ($\eta$) of the diffractive processor is defined as:

$$\eta = \frac{\sum_{n=1}^{N_o}|o'[n]|^2}{\sum_{n=1}^{N_i}|i[n]|^2} \quad (19)$$

**Training of a vaccinated diffractive polarization transformer**

Potential errors caused by fabrication or mechanical assembly imperfections were taken into account by "vaccinating" our model with deliberate random shifts during the training process[57]. Specifically, a random lateral displacement $(D_x, D_y)$ was added to each diffractive layer, where $D_x$ and $D_y$ were independent and identically distributed random variables with a uniform distribution ($U$):

$$D_x \sim U[-0.5\lambda, 0.5\lambda] \quad (20)$$

$$D_y \sim U[-0.5\lambda, 0.5\lambda] \quad (21)$$

A random axial displacement $D_z$ was also added to the axial separations between any two consecutive planes. We generated 6 random variables $\{X_1, X_2, \ldots, X_6\}$, and each of them independently had a uniform distribution $\mathbf{U}[-\lambda, \lambda]$. Then we assigned $D_{z0} = X_1$, and $D_{zi} = X_{i+1} - X_i$ for $i$ = 1, 3, …, 5. Since there were 7 optical elements (including the input and output planes) within the diffractive processor volume, there are a total of 6 free-space propagation processes. The axial distances of these 6 free space propagations from the input plane to the output plane (shown in **Fig. 6b**) were set to $d_0 + D_{z0}$, $d_{p1} + D_{z1}$, $d_1 - d_{p1} + D_{z2}$, $d_{p2} + D_{z3}$, $d_2 - d_{p2} + D_{z4}$, $d_3 + D_{z5}$, respectively.

**Details of the diffractive polarization transformer design used for experimental validation**

Our diffractive polarization transformer design used in the experimental validation consisted of three phase-only diffractive layers ($L_1$-$L_3$) and two polarizer arrays ($PA_1$ and $PA_2$), as shown in the inset of **Fig. 6a**.

Here, each diffractive layer was composed of 80×80 diffractive features, each with a lateral size of 0.6×0.6 $mm^2$. The thickness values of the diffractive features constitute the only trainable parameters within the diffractive volume, which are defined using the following formula in our forward model:

$$h_m(x_m, y_m, z_k) = \frac{h_{\max}}{2} \cdot \left(\sin\left(h'_m(x_m, y_m, z_k)\right) + 1\right) + h_{\text{base}} \tag{22}$$

where $h'_m$ represents the dummy trainable variables. The actual height of each diffractive feature $h_m(x, y)$ was calculated with a maximum allowed thickness of $h_{\max}$ = 1.1 mm and a fixed base thickness of $h_{\text{base}}$ = 0.5 mm.

The amplitude and phase coefficients $a^k(x_m, y_m, z_k)$ and $\phi^k(x_m, y_m, z_k)$ of a diffractive feature can be written as function of $h_m(x_m, y_m, z_k)$ and the incident wavelength $\lambda$:

$$a^k(x_m, y_m, z_k) = \exp\left(-\kappa(\lambda)\frac{2\pi h_m(x_m, y_m, z_k)}{\lambda}\right) \tag{23}$$

$$\phi^k(x_m, y_m, z_k) = (n(\lambda) - n_{\text{air}})\frac{2\pi h_m(x_m, y_m, z_k)}{\lambda} \tag{24}$$

where $n(\lambda)$ and $\kappa(\lambda)$ denote the refractive index and extinction coefficient of the fabrication material, respectively, which together form the complex refractive index $\tilde{n}(\lambda) = n(\lambda) + j\kappa(\lambda)$.

The overall size of a pre-determined polarizer array was set as 40×40 $mm^2$, wherein each linear polarizer has a size of 5×5 $mm^2$. We designed the layout of the linear polarizers within the two polarizer arrays differently, where $PA_2$ is the 180° rotated version of the $PA_1$, i.e., the polarization directions at the same lateral position of two polarizer arrays were set to be orthogonal. Within the input plane, each pixel (aperture) has a size of 12×12 $mm^2$, and the center-to-center distance between two pixels was set as 24 mm, as shown in **Supplementary Fig. S6a**. The output plane has a similar layout as the input plane except that the pixel size was 4×4 $mm^2$ and the center-to-center distance was set to 8 mm, as shown in **Supplementary Fig. S6b**. The centers of the diffractive layers, the input and output apertures, and the polarizer arrays were all aligned to be coaxial using custom-designed holders. The axial distances between the successive layers (including the diffractive layers, polarizer arrays and input/output planes) in our experimental diffractive design were all empirically set to be 25 mm (~33.3$\lambda$). Therefore, in **Fig. 6b** we have $d_0 = d_3 = d_{p1} = d_{p2}$ = 25 mm, and $d_1 = d_2$ = 50 mm.

The fabrication procedures of the THz linear polarizers and polarizer arrays used in our diffractive polarization transformer design follow a photolithography-based approach employing the evaporation of Al on a fused silica ($SiO_2$) substrate[84]. After ultrasonic cleaning in acetone, isopropyl alcohol, and deionized water, photoresist (AZ nLOF 2020, MicroChemicals) was spun coated on a fused silica wafer (JGS2, MSE supplies) serving as the substrate. This coated substrate was then exposed to a UV lamp and developed in developer (AZ MIF 300), followed by evaporating a 700-nm-thickness layer of Aluminum and the liftoff process. As a result of this fabrication process, the Al coating forms patterns of periodic gratings that have a linewidth of ~8 μm and a periodicity of ~20 μm, resulting in a high-performance THz wire-grid polarizer with an experimentally characterized PER of ~30 dB at 0.75 mm (0.4 THz). Photos revealing the microscopic structure of the fabricated THz polarizer array are provided in **Fig. 6c**.

The diffractive layers (see **Fig. 6d**) were fabricated using a 3D printer (Objet30 Pro, Stratasys). The input and output apertures were also 3D printed (Objet30 Pro, Stratasys) and covered with aluminum foil to define the transmission areas. Finally, we used a 3D printed holder (Objet30 Pro, Stratasys) to assemble the fabricated diffractive layers, polarizer arrays and the apertures according to their relative positions set in the forward model of our experimental diffractive design.

Each measurement for the characterization of polarization transformations of our experimental diffractive design was performed by illuminating a single input pixel and employing a specific (x or y) polarization using the corresponding polarizer and the analyzer at the input and output planes, respectively. For such a measurement, we manually blocked three input pixels and left the desired one open to create the illumination. For measuring the signal at the output plane using x or y polarization, another THz linear polarizer was used as the analyzer, positioned right before the output apertures, by rotating its orientation to the desired polarization direction. During these measurements, we scanned the THz detector across all the four pixels (apertures) at the output plane, which represents the measurement of a single row within the submatrices of $\tilde{\boldsymbol{S}}'$ (i.e., $\tilde{\boldsymbol{S}}'^{(\mathrm{exp})}_{\mathrm{xx}}$, $\tilde{\boldsymbol{S}}'^{(\mathrm{exp})}_{\mathrm{xy}}$, $\tilde{\boldsymbol{S}}'^{(\mathrm{exp})}_{\mathrm{yx}}$ and $\tilde{\boldsymbol{S}}'^{(\mathrm{exp})}_{\mathrm{yy}}$) shown in **Fig. 7a, right**. To measure the entire matrix $\tilde{\boldsymbol{S}}'$ of the diffractive polarization transformer, a total of 16 such measurements were acquired, which correspond to all the 2×2×4=16 different combinations of the 2 input polarization components, the 2 output polarization components and the 4 input pixels. In addition to these $\tilde{\boldsymbol{S}}'$ measurements (reported in **Fig. 7a**), we also measured the output polarization fields of our experimental diffractive processor by illuminating input fields with more complex polarization distributions, including spatially varying polarization fields, producing the experimental results reported in **Fig. 7b.**

### Experimental terahertz imaging set-up

As shown in **Fig. 6b**, a terahertz continuous wave (CW) scanning system was used to test our diffractive polarization transformer experimentally. We employed a modular amplifier (Virginia Diode Inc. WR9.0M SGX)/multiplier chain (Virginia Diode Inc. WR4.3x2 WR2.2x2) (AMC) with a compatible diagonal horn antenna (Virginia Diode Inc. WR2.2) as the THz source. A 10 dBm RF input signal at 11.1111 GHz ($f_{RF1}$) was fed into the input of the AMC, which was then multiplied 36 times to produce CW radiation at 0.4 THz. The AMC was also modulated with a 1kHz square wave for lock-in detection. The object plane of the 3D-printed diffractive polarization transformer was placed about 75 cm away from the exit aperture of the horn antenna, resulting in an approximately uniform plane wave incident on its input FOV with a size of 36 mm × 36 mm. After passing through the network, the output signal was 2D scanned with an 8-mm step using a single-pixel mixer (Virginia Diode Inc. WRI 2.2) placed on an XY positioning stage. The positioning stage was built by combining two linear motorized stages (Thorlabs NRT100). A 10 dBm RF signal at 11.0833 GHz (f RF2) was sent to the detector as a local oscillator to down-convert the signal to 1 GHz for further measurement. The down-converted signal was then amplified by a low-noise amplifier (Mini-Circuits ZRL-1150-LN+) and filtered by a 1 GHz (+/-10 MHz) bandpass filter (KL Electronics 3C40-1000/T10-O/O). The signal first passed through a low-noise power detector (Mini-Circuits ZX47-60) and then was measured by a lock-in amplifier (Stanford Research SR830) with the 1kHz square wave serving as the reference signal. The lock-in amplifier readings were calibrated into a linear scale.

**Other implementation details**

All the diffractive polarization transformer models used in this work were trained using Python (v3.7.13) and PyTorch (v1.11.0, Meta Platforms Inc.) with 50 epochs. The learning rate, starting from an initial value of 0.001, was set to decay at a rate of 0.5 every 10 epochs. For the training of our diffractive polarization transformer models, we used a workstation with an RTX 3090 graphical processing unit (GPU, Nvidia Inc.), an Intel® Core™ i9-12900F central processing unit (CPU, Intel Inc.) and 64 GB of RAM, running Windows 10 operating system (Microsoft Inc.). The training time of a diffractive polarization transformer with $N = 8N_i N_o$ is ~5 hours.

**Supporting Information:** This file contains training loss function details and **Supplementary Figures S1-S6**.

## Figures

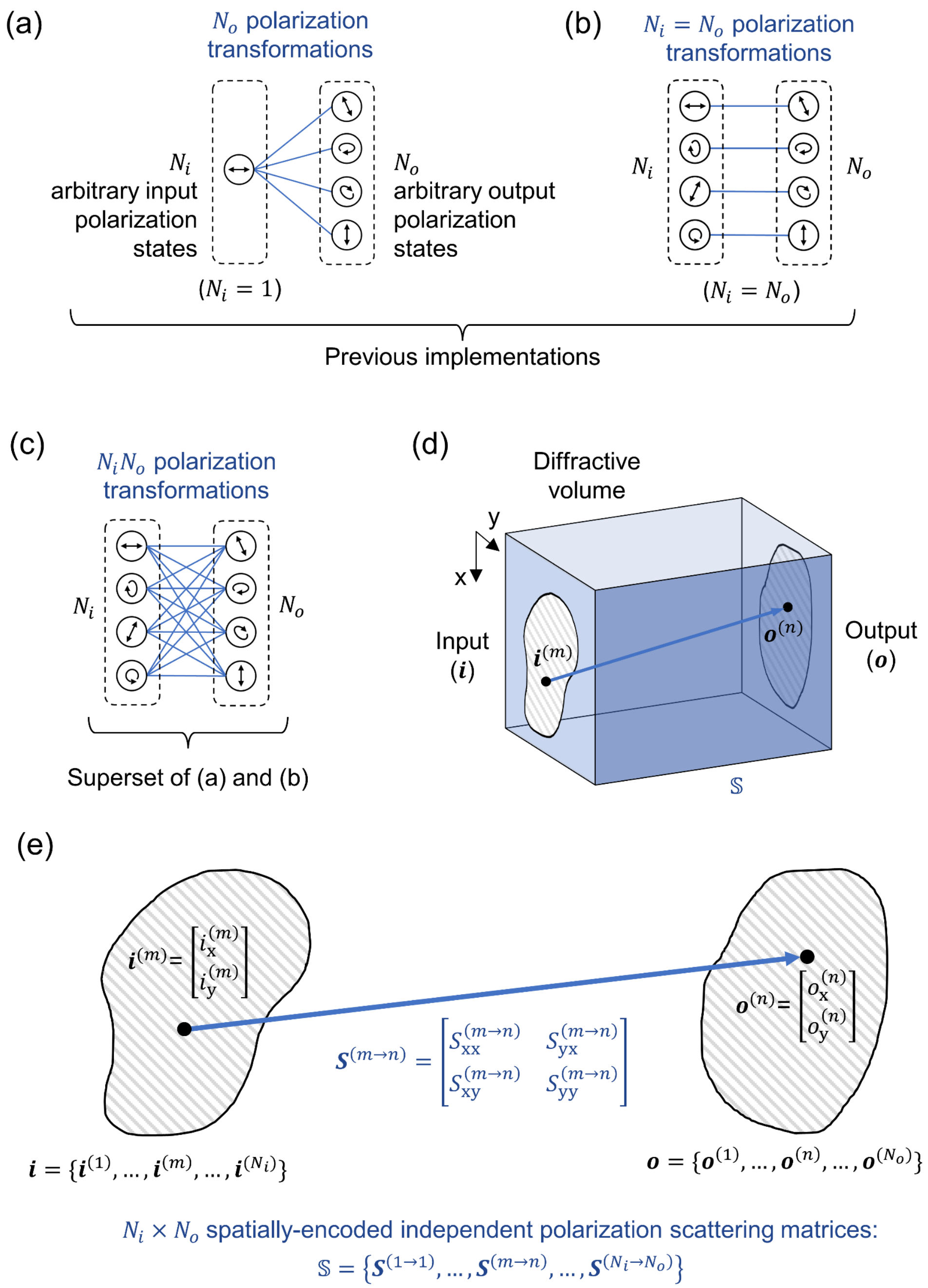

**Fig. 1. Universal polarization transformations between spatially varying polarization fields. a-c**, Conceptual abstraction of different implementations of polarization transformations. Among these implementations, (a) represents transformations from a single ($N_i = 1$) input polarization state to a plurality of ($N_o > 1$) output polarization states, which indicates $N_o$ independent polarization transformations performed based on an input field with a single, uniform polarization state. (b) represents the transformations from multiple ($N_i$) input polarization states to multiple ($N_o$) output polarization states in the form of one-to-one mapping ($N_i = N_o$), indicating $N_i = N_o$ distinct polarization transformations performed at multiple spatial locations in parallel based on a spatially varying input polarization field. (c) represents our work, universally covering all the transformations with a complete mapping from multiple ($N_i$) input polarization states to multiple ($N_o$) output polarization states, which forms $N_i N_o$ independent polarization transformations from one spatially varying polarization field to another, representing the universal polarization transformations achieved by our framework. **d**, The concept of a diffractive volume in 3D space, where an input FOV and an output FOV are connected by a linear, coherent optical system depicted as a black-box. **e**, The relationship between each polarization state at the input FOV ($\boldsymbol{i}^{(m)}$) and each polarization state at the output FOV ($\boldsymbol{o}^{(n)}$) can be depicted through a 2×2 complex-valued polarization scattering matrix $\boldsymbol{S}^{(m\to n)}$. By collecting all the independent $\boldsymbol{S}^{(m\to n)}$ matrices corresponding to all the different combinations of input and output polarization states, we form a set $\mathbb{S} = \{\boldsymbol{S}^{(1\to 1)}, \dots, \boldsymbol{S}^{(m\to n)}, \dots, \boldsymbol{S}^{(N_i\to N_o)}\}$ that contains $N_i N_o$ spatially-encoded polarization scattering matrices, which completely describe the transformation relationship between an input polarization field and an output polarization field that are both spatially varying.

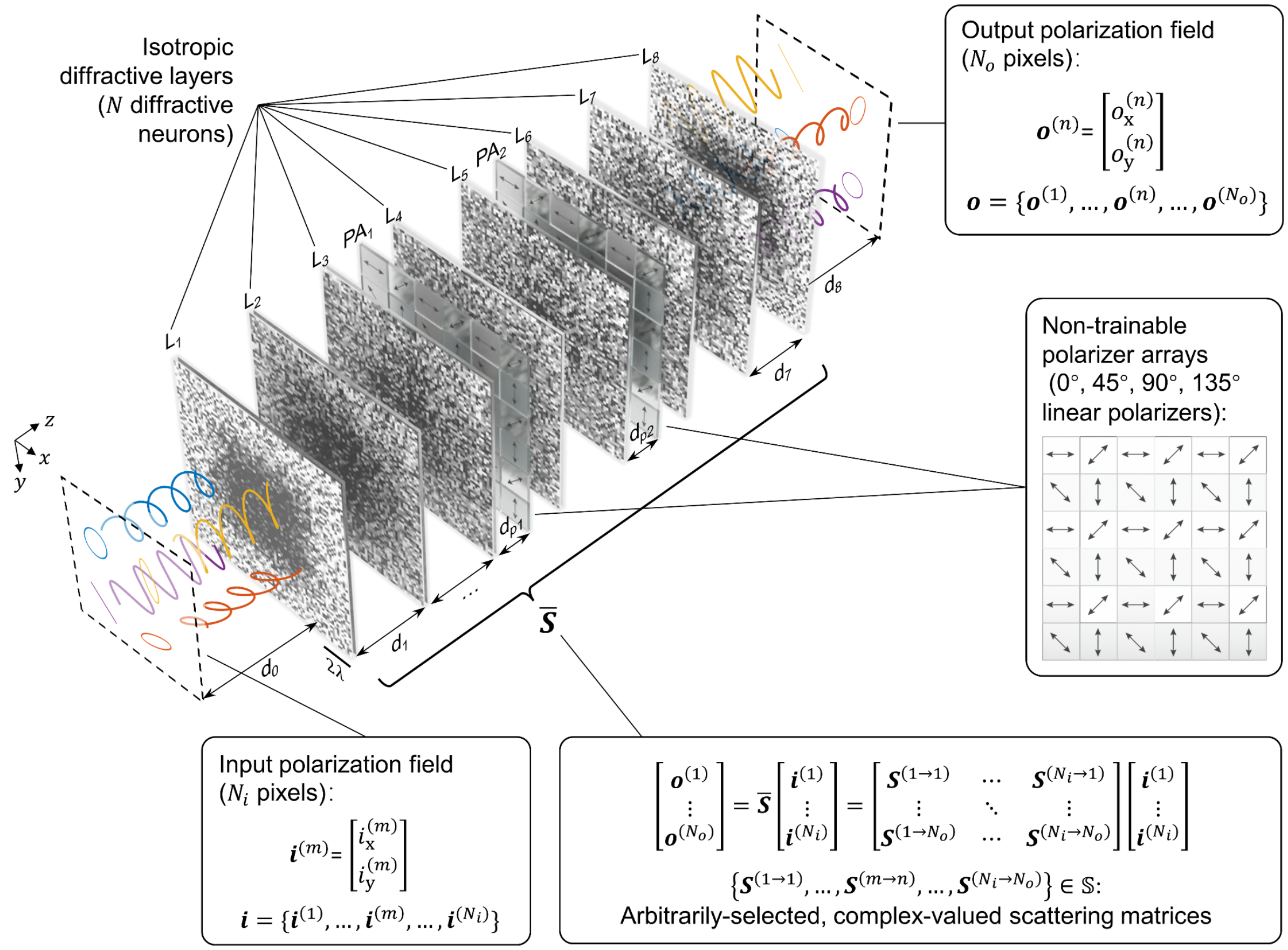

**Fig. 2. Schematic of a diffractive universal polarization transformer.** This diffractive polarization transformer comprises two pre-determined arrays of linear polarizers and eight trainable isotropic diffractive layers, axially covering a total length of ~89.6λ. The polarizer arrays consist of a plurality of square linear polarizers, which have polarizing orientations selected from 0°, 45°, 90°, and 135°, forming multiple 2×2 patterns that are periodically arranged along the x and y directions. After the training of the transmission coefficients of the diffractive features on the eight isotopic diffractive layers, this diffractive polarization transformer can all-optically synthesize a large set ($\mathbb{S}$) composed of $N_i N_o$ spatially-encoded polarization scattering matrices (i.e., $\mathbb{S} = \{\boldsymbol{S}^{(1\to1)}, \ldots, \boldsymbol{S}^{(m\to n)}, \ldots, \boldsymbol{S}^{(N_i\to N_o)}\}$) between a polarized input field $\boldsymbol{i}$ and a polarized output field $\boldsymbol{o}$, where $N_i$ and $N_o$ represent the number of useful pixels in the fields $\boldsymbol{i}$ and $\boldsymbol{o}$, respectively. PA: polarizer array.

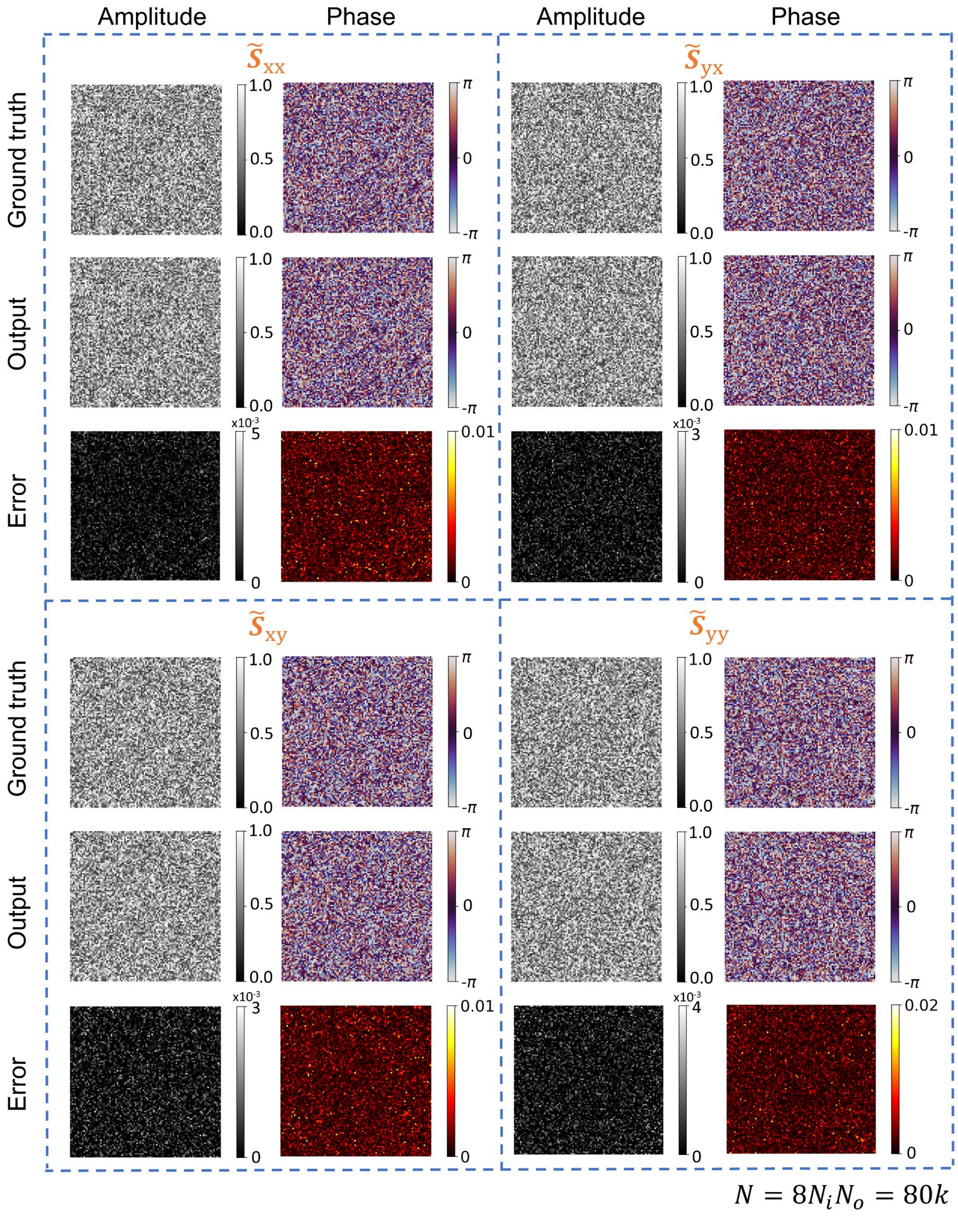

**Fig. 3**. **Amplitude and phase of the ground truth submatrices $\tilde{S}_{xx}, \tilde{S}_{yx}, \tilde{S}_{xy}$ and $\tilde{S}_{yy}$ of the scattering matrix $\tilde{S}$, compared with their corresponding all-optically implemented versions $\tilde{S}'_{xx}, \tilde{S}'_{yx}, \tilde{S}'_{xy}$ and $\tilde{S}'_{yy}$ using the diffractive polarization transformer design shown in Fig. 2 with $N = 8N_iN_o$.** The relative amplitude and phase errors, $\left||\tilde{S}_*| - |\tilde{S}'_*|\right|$ and $\left|\angle\tilde{S}_* - \angle\tilde{S}'_*\right|_\pi$, are also shown, where $|*|_\pi$ indicates the wrapped phase difference.

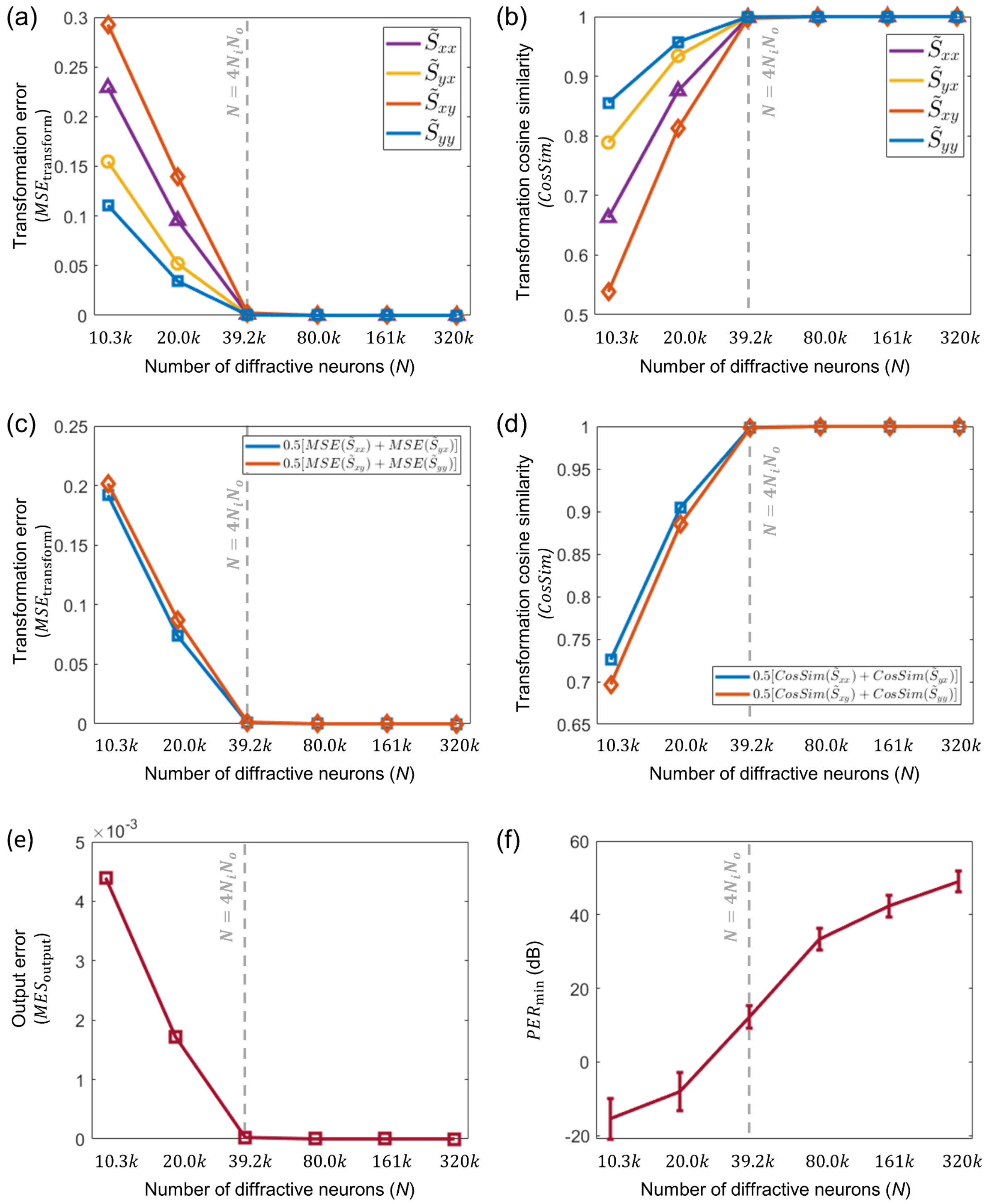

**Fig. 4. Performance evaluation of the diffractive universal polarization transformer designs shown in Fig. 2 as a function of the number of diffractive neurons/features ($N$).** (a) $MSE_{\text{transform}}$, (b) $CosSim$, (e) $MSE_{\text{output}}$ and (f) $PER_{\min}$ are reported as a function of $N$ to evaluate the performance of the diffractive polarization transformer designs. (c) and (d) are derived based on (a) and (b), respectively, by averaging the performance metrics ($MSE_{\text{transform}}$ or $CosSim$) quantified for the submatrices corresponding to the same polarization states (x or y) at the output.

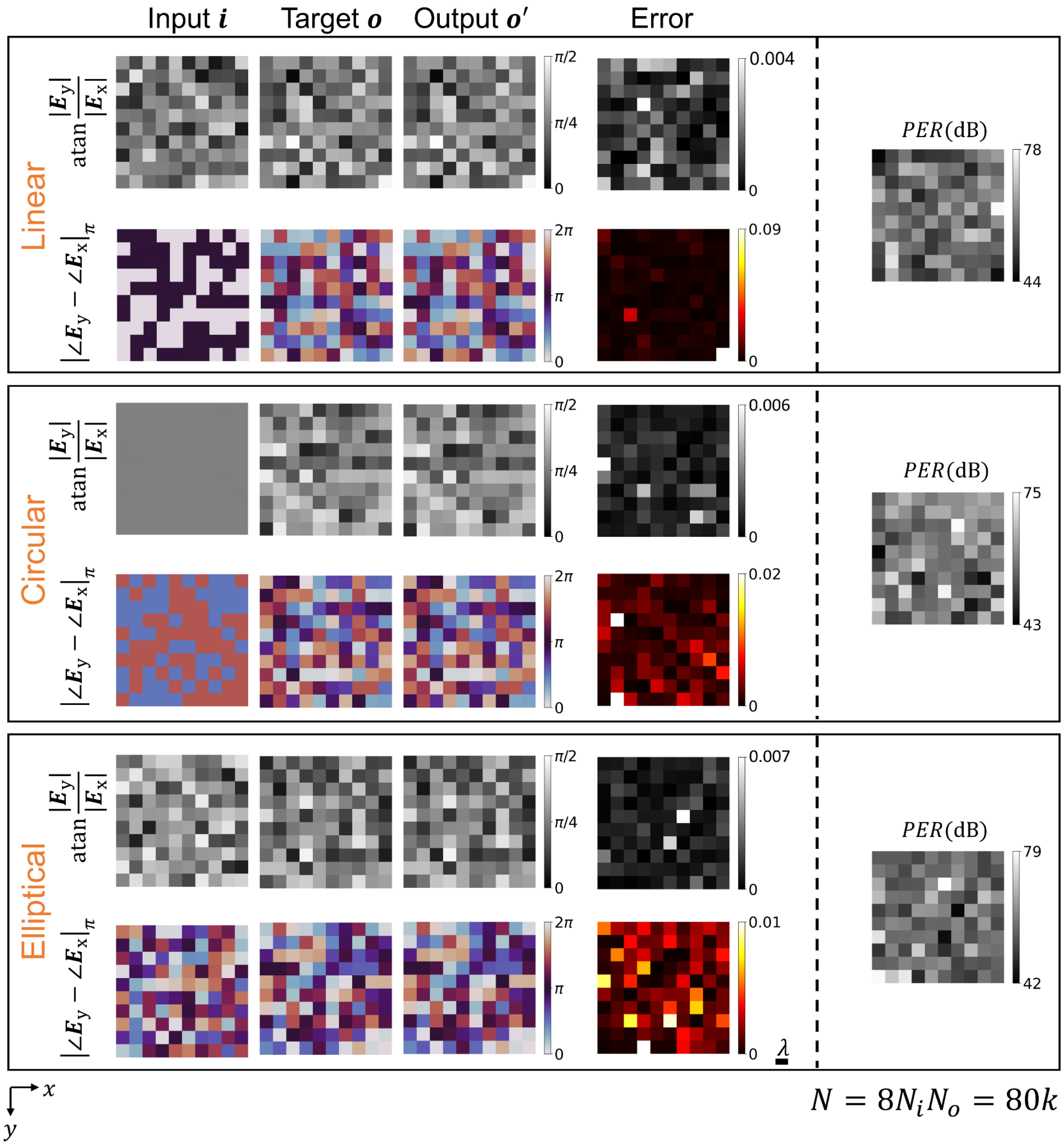

**Fig. 5. Examples of the input and ground truth (target) output polarization fields corresponding to the target polarization transformations in Fig. 3, along with the output fields resulting from our diffractive polarization transformer design shown in Fig. 2 ($\boldsymbol{N = 8N_iN_o}$).** For these polarization fields, $\text{atan}\frac{|E_y|}{|E_x|}$ and $\left|E_y - E_x\right|_\pi$, are calculated to evaluate the error between the target fields and the all-optical output fields. $PER$ is also shown to compare the power of the field component with the desired polarization state to its counterpart with the undesired (orthogonal) polarization state. Three different types of polarization were used for generating the input polarized fields, including linear only (top), circular only (middle) and elliptical polarization (bottom). $|*|_\pi$ indicates the wrapped phase difference.

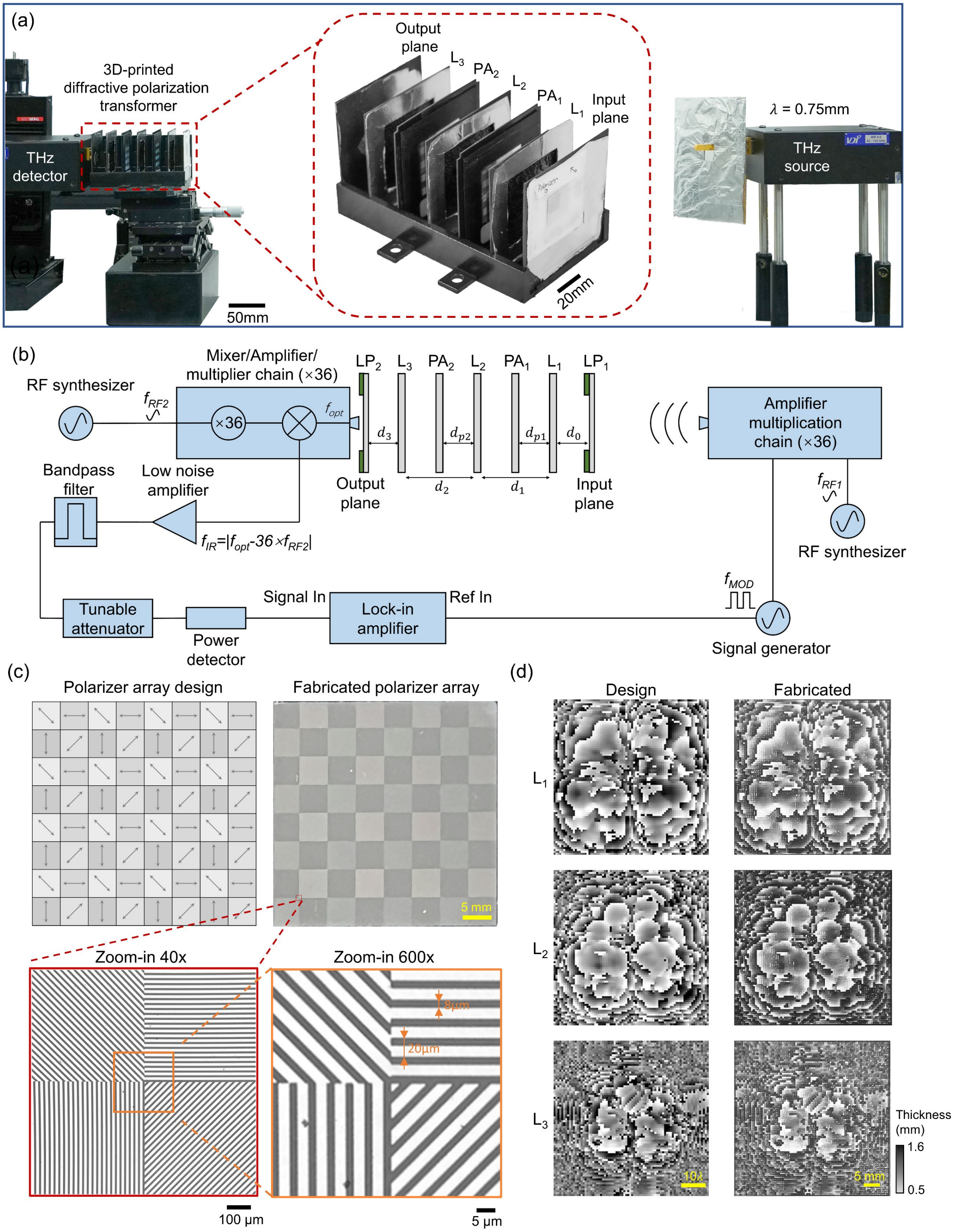

**Fig. 6. Experimental set-up of the diffractive universal polarization transformer. a**, Photograph of the experimental set-up, including the fabricated diffractive polarization transformer. **b**, Schematic of the THz

set-up. **c,** Designed layout of the polarizer array (top left) and the photograph of its fabricated version (top right), along with the microscopic images revealing the wire-grid structures of the fabricated polarizer array, imaged using an optical microscope operating in transmission mode (bottom). The orientations of the linear polarizers in this polarizer array are identical to those used in the polarizer array $PA_1$, while the orientations of the linear polarizers in $PA_2$ are 180-degree rotated versions of those in $PA_1$. LP: linear polarizer. **d**, Learned thickness profiles of the diffractive layers (left) and the photographs of their fabricated versions using 3D printing (right).

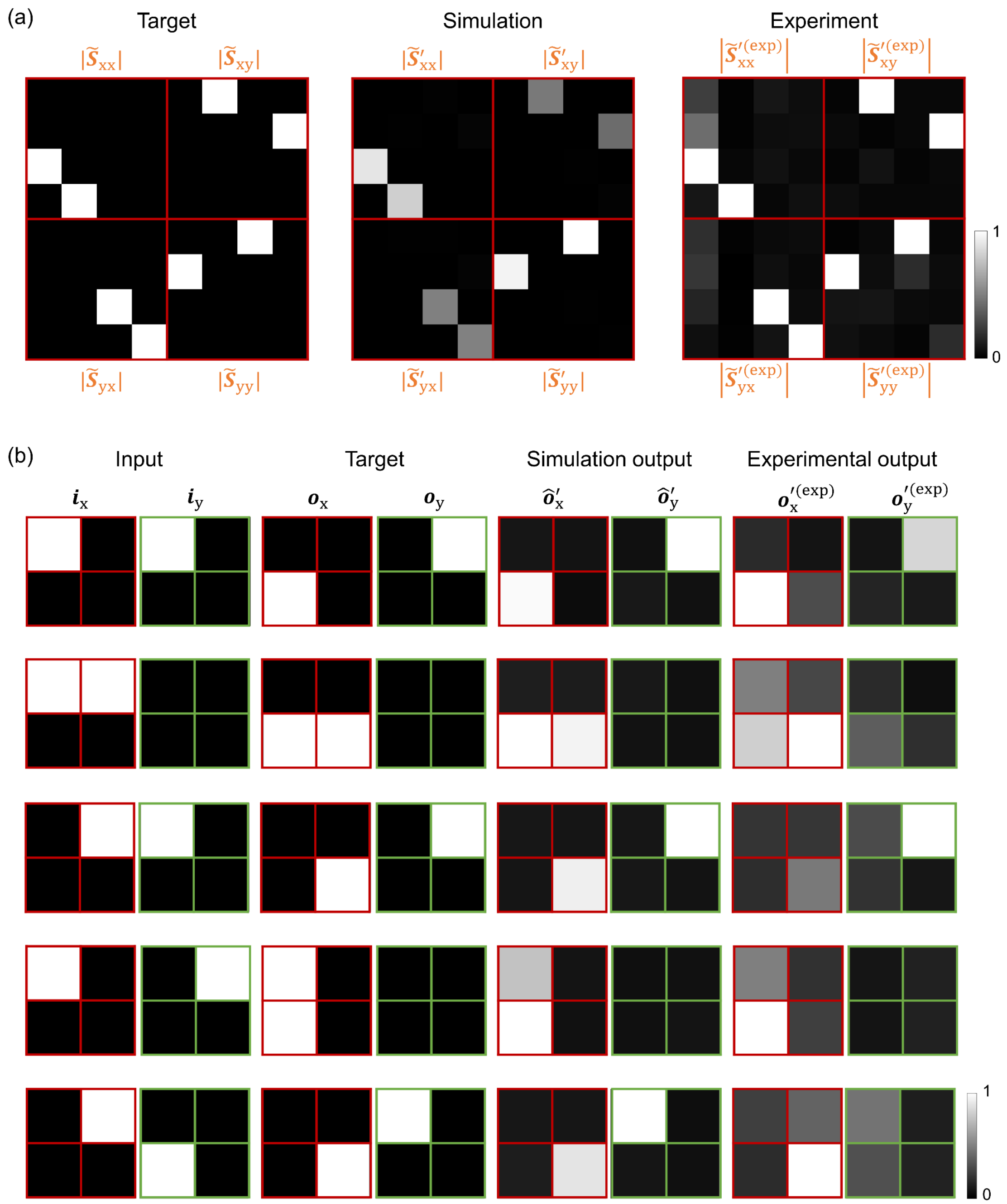

**Fig. 7. Experimental results of the diffractive universal polarization transformer for all-optical polarization permutation operation. a**, The experimental characterization of the all-optical polarization transformations implemented by our diffractive polarization transformer reveals a good agreement with its numerical counterparts and the ground truth. **b**, The experimental measurements of the diffractive output polarization fields when using input fields with spatially varying polarization states show minimal differences compared to their numerical counterparts and the ground truth.